\documentclass[]{raa}

\usepackage{graphicx,times}
\usepackage{natbib}
\usepackage{subfigure}
\providecommand{\bm}[1]{\mbox{\boldmath$#1$\unboldmath}}

\newcommand\rat{\ifmmode{\cal R}\else{${\cal R}$}\fi}
\newcommand\CO{$^{12}$CO}

\begin{document}

\title{$^{12}$CO, $^{13}$CO and C$^{18}$O observations along the major axes of nearby
bright infrared galaxies}
\volnopage{{\bf XXXX} Vol.\ {\bf X} No. {\bf XX}, 000--000}
\setcounter{page}{1}

\author{Q.H. Tan \inst{1,2,3}
 \and Yu Gao \inst{1}
 \and Z.Y. Zhang \inst{1,2}
 \and X.Y. Xia \inst{3}}

\institute{Purple Mountain Observatory, Chinese Academy of Sciences,
Nanjing 210008, China \\
{\it qhtan@pmo.ac.cn}
 \and
      Graduate School of Chinese Academy of Sciences, Beijing 100039, China
   \and
      Center of Astrophysics, Tianjin Normal University, Tianjin 300384, China
 \vs \no
{\small Received [0] [0] [0]; accepted [0] [0] [0]}}

\abstract{We present simultaneous observations of\ \CO,
$^{13}$CO and C$^{18}$O $J$=1$-$0 emission in 11
nearby ($cz<$1000\,km s$^{-1}$) bright infrared galaxies. Both \CO\
and $^{13}$CO are detected in the centers of all galaxies, except
for $^{13}$CO in NGC~3031. We have also detected C$^{18}$O,
CS$J$=2$-$1, and HCO$^+ J$=1$-$0 emission in the nuclear regions of
M82 and M51. These are the first systematical extragalactic
detections of \CO\ and its isotopes from the PMO 14m telescope.\\
We have conducted half-beam spacing mapping of M82 over an area of
$4'\times 2.5'$ and major axis mapping of NGC~3627, NGC~3628,
NGC~4631, and M51. The radial distributions of \CO\ and $^{13}$CO in
NGC~3627, NGC~3628, and M51 can be well fitted by an exponential
profile. The \CO/$^{13}$CO intensity ratio, \rat,\ decreases
monotonically with galactocentric radius in all mapped sources. The
average \rat \ in the center and disk of the galaxies are
9.9$\pm$3.0 and 5.6$\pm$1.9 respectively, much lower than the
peculiar \rat($\sim$24) found in the center of M82.\\
The intensity ratios of {$^{13}$CO/C$^{18}$O}, {$^{13}$CO/HCO$^+$}
and {$^{13}$CO/CS} (either ours or literature data) show little variations
with galactocentric radius, in sharp contrast with the greatly
varied \rat. This supports the notion that the observed gradient in
\rat \ could be the results of the variations of the physical
conditions across the disks. The H$_2$ column density derived from
C$^{18}$O shows that the Galactic standard conversion factor
($X$-factor) overestimates the amount of the molecular gas in M82 by
a factor of $\sim$2.5. These observations suggest that the
$X$-factor in active star-forming regions (i.e., nuclear regions)
should be lower than that in normal star-forming disks, and the
gradient in $\rat$ can be used to trace the variations of the
$X$-factor.
}

\authorrunning{Tan et al.}
\titlerunning{$^{12}$CO, $^{13}$CO and C$^{18}$O observations in nearby galaxies}
\maketitle

\section{Introduction}
\label{sect:intro}

It is well known that molecules constitute a significant fraction of
the interstellar medium (ISM) and stars form from molecular clouds.
However, the most abundant molecule, H$_2$, cannot be detected
directly in typical cold (10-40\,K) molecular cloud, owing to its
lack of a permanent electric dipole moment. The next most abundant
molecule is  \CO, which is found to be an excellent tracer of H$_2$
due to the low excitation temperature ($\sim$10\,K) and critical
density ($\sim$300\,cm$^{-3}$) (\citealt{Evans99}). Generally, the lowest
transition ($J$=1$-$0) of the rotational lines of \CO\ and its optically
thin isotopic variants (e.g., $^{13}$CO and C$^{18}$O) can be used
to estimate the column density ($N$(H$_2$)) of molecular clouds in
our Galaxy (\citealt{Frerking82, Young82, Wilson09}). However, up to now,
the column density of molecular gas
in galaxies is difficult to obtain directly in this way due to the
beam dilution and the weakness of line emission in CO isotopes.
Accurate determination of the H$_2$ column densities from CO
observations has therefore been a longstanding challenge in external
galaxies.

In our Galaxy, results from independent methods show a tight
correlation between the integrated \CO\ line intensity $I_{\rm ^{12}CO}$
and $N$(H$_2$), and the ratios of $N$(H$_2$) to $I_{\rm ^{12}CO}$ appear
to be constant with values of (1-5)$\times$10$^{20}$\,cm$^{-2}{\rm
K}^{-1}{\rm km}^{-1}$s across the Galaxy (\citealt{Bohlin78, Dickman78,
Hunter97, Scoville87, Young91}). This constant value is denoted as the
\CO-to-H$_2$ conversion factor, or the $X$-factor~($X\equiv N({\rm H}_2)/I_{\rm
^{12}CO}$). It is beyond the scope of this paper to analyze in great
detail the origin of the empirical '$X$-factor' since many studies
have shown that the $X$-factor varies with different physical
conditions and environments, and is influenced by the CO abundance,
the gas excitation and the radiative transfer processes (\citealt{Scoville74}).
For example, the amounts of molecular gas in the
metal-poor Magellanic Clouds and M31 were claimed to be
underestimated by up to a factor of $\sim$10 if a standard Galactic
$X$-factor was adopted (\citealt{Maloney88, Allen93, Israel97}).
Nevertheless, recent imaging of CO clouds
reports similar standard $X$-factor in metal-poor galaxies (\citealt{Bolatto08}).
Conversely, the amounts of molecular gas in the
ultraluminous infrared galaxies (ULIRGs) might be overestimated by
factors of $\sim$5 (\citealt{Downes98}) using the standard
Galactic $X$-factor.

The $IRAS$ all-sky survey has revealed a large population of infrared bright
galaxies with bulk energy emission in the far-infrared (\citealt{Soifer87, Sanders03}),
which is mostly due to dust heating from starburst (\citealt{Sanders96}).
Early numerous CO observations of infrared galaxies found that these galaxies are
rich in molecular gas (\citealt{Young95}), and there exist
a correlation between CO and far-infrared luminosity (\citealt{Solomon88, Young91}),
though the correlation is non-linear (\citealt{Gao04b}) since $L_{\rm IR}/L_{\rm CO}$
correlates with $L_{\rm IR}$.
Moreover, recent studies on high-redshift star forming galaxies further confirmed the validity
of CO-IR luminosity correlation (\citealt{Solomon05, Daddi10}).

Although large quantities of observational studies have aimed at mapping the
molecular gas distribution/kinematics in nearby infrared bright galaxies with
single-dish and/or interferometer telescopes (\citealt{Braine93, Young95,
Sakamoto99, Nishiyama01, Helfer03, Leroy09}), these surveys are
almost all based on the observations of CO $J$=1$-$0 or $J$=2$-$1,
and only limited systematical studies of CO isotope variants such as
$^{13}$CO, have been published so far (\citealt{Paglione01}). Owing
to the large isotope abundance ratio [\CO]/[$^{13}$CO]$\approx$30-70
across the Galaxy (\citealt{Langer90}), the opacity of $^{13}$CO
is much lower than that of \CO, and thus $^{13}$CO is believed to be
mostly optical thin and trace the molecular gas column density
adequately in most cases. Consequently, the variations in the
intensity ratio of \CO\ to $^{13}$CO as a function of galactocentric
radius, could give a reliable test of whether $X$-factor varies
systematically within galaxies. Studies of the Galactic $X$-factor
have revealed large variations in its value towards the Galactic
nuclear region, with an order of magnitude increase from the center
to the outer disk (\citealt{Sodroski95, Dahmen98}). In
other galaxies, however, \cite*{Rickard85} claimed that the
value of integrated intensity ratio of \CO\ to $^{13}$CO in the
nucleus is on average a factor of two higher than that in the disk. In
fact, both \cite*{Young86} and \cite*{Sage&Isbell91} suggested that there
was no clear evidence of a systematic difference in \rat \ within a galaxy,
and they suggested that the
variations in \rat \ observed by \cite*{Rickard85} is likely
to be caused by pointing errors. Recently, mapping surveys of \CO\ and
$^{13}$CO emission towards a large sample of nearby galaxies were
carried out by \cite*{Paglione01}, finding that the same physical
processes (e.g., the molecular gas kinetic temperature) may both
affect \rat \ and $X$-factor, moreover, the latter is also expected
to decrease from the disks to the nuclear regions by factors of 2-5.

However, any study of the physical properties of molecular gas that involves
using line intensity ratios might be influenced by measurement
errors, owing to the telescopes' different beam sizes, uncertain beam
efficiencies, pointing errors, and calibrations. Recently updated and
improved sensitivity and system performance with the Purple Mountain
Observatory~(PMO) 14m telescope at Delingha, China
allow us to simultaneously observe extragalactic sources systematically
in \CO, $^{13}$CO, and C$^{18}$O for the first time. Consequently,
our simultaneous observations of three CO isotope variants with same
telescope are better suited to obtain the well-calibrated line intensity
ratios than observations from different telescopes carried out
with different tunings at different time.

In this paper, we present the results of simultaneous observations
of \CO, $^{13}$CO, and C$^{18}$O along the major axes in nearby
infrared bright galaxies. The sample selection, observations, and
data reduction are described in \S~\ref{sect:sample}; the observed
CO spectra, CO radial distributions, and position-velocity diagrams
are presented in \S~\ref{sect:results}. These results together with
the radial distributions of molecular gas, CO isotopic ratio \rat,
and possible physical mechanisms that could be responsible to the
observed variations in \rat \ are discussed in \S~\ref{sect:discussion}.
Finally a summary is presented in \S~\ref{sect:summary}. A stability
analysis of the PMO 14m telescope is presented in Appendix
~\ref{stability}.

 \section{Sample, Observations and Data Reduction}
\label{sect:sample}

The galaxies in this study were selected from the Revised Bright
Galaxy Sample \citep[RBGS,][]{Sanders03}. We selected 11 galaxies
based on the following three criteria: (1) $f_{60\mu m} \geq$ 50\,Jy
or $f_{100\mu m} \geq$ 100\,Jy. This infrared flux cutoff was chosen
to ensure both \CO\ and $^{13}$CO could be detected with reasonable
integration time since it is well known that the infrared luminosity
of galaxies is correlated with the CO luminosity \citep[e.g.,][]{Solomon88}.
(2) $cz \leq$ 1000\,km s$^{-1}$. This velocity limit was
chosen due to the limited tuning range of the SIS receiver with the
PMO 14m telescope. (3)9h$\leq$ R.A. $\leq$15h and
Decl.$\geq0^{\circ}$. For the reason that we can take full advantage
of the Galactic dead time to observe galaxies in the northern sky.
Some general properties of the sample galaxies are summarized in
Table~\ref{Table1}.

Our observations were made between 2008 January and 2009 December
using the PMO 14m millimeter telescope at the Delingha.
We used the
3mm SIS receiver operated in double-sideband mode, which allowed for
simultaneous observations of three CO isotope variants, with \CO\ in
the upper sideband and $^{13}$CO and C$^{18}$O in the lower
sideband. The Half Power Beam Width (HPBW) is $\sim$ 60$''$, and the
main beam efficiency $\eta_{\rm mb}$ = 0.67. Typical system
temperatures during our runs were about 180-250 K. The FFT
spectrometer was used as back end with a usable bandwidth of 1 GHz
and a velocity resolution of 0.16\,$\rm{km~s}^{-1}$ at 115GHz.
Observations were done in position switching mode and calibrated
using the standard chopper wheel method. The absolute pointing
uncertainty was estimated to be $\sim$10$''$ from continuum
observations of planets, and the pointing were checked every two
hours by taking spectra toward the standard source IRC+10216
throughout our observations. Each galaxy was first observed at the
center position, and then along its major axis from the center to
the outer disk positions separated by half-beam size. Besides the CO
observations, we also observed the dense molecular gas tracers
HCO$^+$ and CS in the nuclear regions of galaxies.

The data were reduced using CLASS, which is part of the
GILDAS\footnote{http://iram.fr/IRAMFR/GILDAS/}
software package. All scans were individually inspected, and those
with either strongly distorted baselines or abnormal rms noise
levels were discarded. Line-free channels which exhibited positive
or negative spikes more than 3 $\sigma$ above the rms noise were
blanked and substituted with values interpolated from the adjacent
channels, and then a linear baseline was subtracted from the
'line-free' spectrum. After correcting each spectrum for main beam
efficiency $\eta_{\rm mb}$, the temperature scale of the spectra was
converted to the main beam temperature $T_{\rm mb}$ scale from the
$T^*_A$. The spectra were then co-added, weighted by the inverse of
their rms noises, and the final spectral resolution was smoothed to
$\sim$20\,$\rm{km~s}^{-1}$.

Since this is the first time that the PMO 14m millimeter telescope
systematically observed galaxies other than our own, relatively
rather long observing time at each pointing is devoted to accumulate
relatively adequate integration time. Thus our observations are the
first long integrations for the PMO 14m telescope ever conducted and
offer an excellent opportunity to test the stability of the upgraded
system (see Appendix ~\ref{stability}). The work present here
represents the total of $\sim$ 500 hours observing time in CO observations
before the telescope system maintenance in the summer of 2009 at the
PMO 14m and $\sim$ 200 hours after the upgrade, with $\sim40\%$ data
discarded owing to the unstable spectrum baseline and bad weather conditions.

\section{Results and analysis}
\label{sect:results}
\subsection{CO Isotopic Spectra}
\label{ss:spectra}

We detected \CO\ emission from the centers of all 11 galaxies
observed, of which 10 were also detected in $^{13}$CO
simultaneously. Four galaxies (NGC~3627, NGC~3628, NGC~4631 and M51)
were observed along the major axes, and \CO\ emission was detected at
all 42 off-center positions along the major axes, while $^{13}$CO
was detected at 27 out of the 42 positions. For M82, the starburst
galaxy, \CO\ emission was detected at all 47 positions that were
mapped in the central $4'\times 2.5'$ region, while $^{13}$CO was
tentatively detected at 15 positions. C$^{18}$O emission was only
detected at 13 points close to the nuclear regions in M51 and M82.

Here, we focus on presenting $^{13}$CO and C$^{18}$O spectra as
compared to \CO\ observed simultaneously at the same positions, \CO\
spectra at those $^{13}$CO undetected positions won't be shown here.
Since many previous observations of \CO\ emission in these galaxies
are available in literature \citep[e.g.,][]{Young95}, and our \CO\
observations show similar results and comparable spectra. The
spectra of both \CO\ and $^{13}$CO, as well as {\sl Spitzer} IRAC
3.6~$\mu$m, 8~$\mu$m and MIPS 24~$\mu$m infrared composite color
images showing the observed beam positions of the mapped galaxies,
are shown in Fig.1. All the three CO isotopic spectra at
these C$^{18}$O detected positions are shown in Fig.~\ref{Fig2}.

\subsection{Derived Parameters}
\label{ss:parameters}

The observational results and derived properties are summarized in
Table~\ref{Table2}. The velocity-integrated \CO\ (and isotopes)
intensity, $I_{\rm CO}\equiv \int T_{\rm mb}dv$, is obtained by
integrating $T_{\rm mb}$ over the velocity range with CO line
emission feature. Using the standard error formulae in \cite*{Gao96}
\cite*[also in][]{Matthew01}, the error in the integrated intensity is
\begin{equation}
\Delta I = T_{\rm rms} \ \Delta v_{\rm FWZI} / [ f \ (1-\Delta
v_{\rm FWZI}/W)]^{1/2} \ [{\rm K~km~s}^{-1}],
\end{equation}
where $T_{\rm rms}$ is the rms noise in the final spectrum in mK,
$f =\Delta v_{\rm FWZI} / \delta_v$, where $\Delta v_{\rm FWZI}$ is the
linewidth of the emission feature,  $\delta_v$ is the channel bin
spacing, and $W$ is the entire velocity coverage of the spectrum in
units of kilometers per second. For non-detections (only CO
isotopes), some spectra are further smoothed and found to be
marginally detected with signal-to-noise of $\sim$2.5. Otherwise, a
2 $\sigma_{I}$ upper limits are given based on the estimation by using
the expected line width from the detected \CO\ lines at exactly the
same position.

The H$_2$ column density and mass surface density for galaxies in
our sample are derived from the empirical relations (\citealt{Nishiyama01})
 \begin{equation}
N({\rm H}_2)[{\rm cm}^{-2}] = 2\times10^{20}I_{\rm
^{12}CO}[{\rm K~km~s^{-1}}],
\end{equation}
and
 \begin{equation}
     \Sigma({\rm H}_2)[M_\odot{\rm pc}^{-2}] = 3.2I_{\rm ^{12}CO}
[{\rm K~km~s}^{-1}] {\rm cos}i,
\end{equation}
where cos$i$ corrects the mass to face-on and a Galactic \CO-to-H$_2$
conversion factor $X$=2.0$\times$10$^{20}$\,cm$^{-2}{\rm K}^{-1}{\rm
km}^{-1}$s is adopted (\citealt{Dame01}). Obviously, it can be seen
from Table~\ref{Table2} that the column density in M82 is usually a
magnitude higher than that in normal spiral galaxies (the rest of
the sample).

Assuming that $^{13}$CO has the same excitation temperature as \CO\
and the molecular cloud is under LTE conditions, then we can
calculate the average $^{13}$CO optical depth from
 \begin{equation}
\tau(^{13}{\rm CO}) = {\rm ln}[1-\frac{\int T^{*}_R(^{13}{\rm CO})dv}
{\int T^{*}_R(^{12}{\rm CO})dv}]^{-1},
\end{equation}
where $T^{*}_R$ should be corrected for filling factor, and we can
only estimate an average over all of the unresolved clouds in the beam.
Using the definition in \cite*{Wilson09}, the H$_2$ column
density can be derived from the $^{13}$CO line intensity as
 \begin{equation}
N({\rm H_2})(^{13}{\rm CO})=[\frac{\tau(^{13}{\rm CO})}
{1-e^{-\tau(^{13}{\rm CO})}}] 2.25\times10^{20}\frac{\int T_{\rm
mb}(^{13}{\rm CO})dv} {1-e^{-5.29/T_{\rm ex}}},
\end{equation}
where the $^{13}$CO abundance [$^{13}$CO]/[H$_2$] is 8$\times10^{-5}$/60
(\citealt{Frerking82}) and the excitation temperature, $T_{\rm ex}$, is taken
to be the kinetic temperature of the gas, $T_{\rm K}$. Thus, we can
estimate $T_{\rm K}$ by equating the column densities derived from
both \CO\ and $^{13}$CO. Both the derived $\tau(^{13}{\rm CO})$ and
$T_{\rm K}$ are listed in Table~\ref{Table2}. Note that LTE assumption
is most likely invalid in the central regions of M82. Therefore, the extremely
low optical depth of $^{13}$CO ($\sim$0.04) should be treated only as a lower
limit, and the resulting kinetic temperature (between 70 and 120~K),
which are about 3-4 times higher than that in normal galaxies, should be
treated as an upper limit.

Similarly, we estimated the optical depth of C$^{18}$O adopting
the same method as equation (4), and derived the H$_2$ column
density from C$^{18}$O intensity using (\citealt{Sato94})
 \begin{equation}
N({\rm H_2})({\rm C}^{18}{\rm O})=[\frac{\tau(^{18}{\rm CO})}
{1-e^{-\tau(^{18}{\rm CO})}}] 1.57\times10^{21}\frac{\int T_{\rm
mb}({\rm C^{18}O})dv} {1-e^{-5.27/T_{\rm ex}}},
\end{equation}
where the abundance of [C$^{18}$O]/[H$_2$] is
1.7$\times10^{-7}$ (\citealt{Frerking82}) and $T_{\rm ex}$ is
adopted from the value listed in Table~\ref{Table2}. The derived
values of $\tau{\rm (C^{18}O)}$ and $N({\rm H_2})({\rm C}^{18}{\rm
O})$ for the 13 points of M82 and M51 detected in C$^{18}$O are
listed in Table~\ref{Table3}. The average optical depths of
C$^{18}$O in M82 and M51 are 0.02 and 0.05 respectively, both are
about three times lower than that of $^{13}$CO. Therefore, although the optical
depth of $^{13}$CO is moderate ($\tau(^{13}{\rm CO}) \sim$0.3--0.4) in a few galaxies
(e.g., NGC~3627 and NGC~4631; see Table~\ref{Table2}), C$^{18}$O is always optically
thin in all cases here.

\subsection{Distribution of CO Isotopes and Line Ratios}
\label{ss:distribution}

The distributions of \CO\ and $^{13}$CO velocity-integrated
intensities and their ratio \rat \ as a function of galacto-central
radius are shown in Fig.~\ref{Fig3}. Note that none of these
profiles have been corrected for inclination. Both the \CO\ and
$^{13}$CO emission line intensities show an exponential decrease
with radius, and the ratio \rat \ decreases from nucleus to the
outer disk.

\subsubsection{Radial distributions of \CO\ and $^{13}$CO}
\label{sss:radial}

The obvious trends shown in Fig. 3 are that the observed radial
distribution of $^{13}$CO follows that of \CO. The integrated line
intensities usually peak in the center of galaxies and fall
monotonically toward larger galactocentric radii. For the barred
spiral galaxies NGC~3627 and NGC~4631, however, both \CO\ and
$^{13}$CO intensities peak at radius of $\sim$ 1\,kpc ($\sim$0.5')
rather than at the nuclei. The same feature has also been presented
in \cite*{Paglione01} for NGC~4631. Some previous high resolution
observations of barred galaxies revealed that the molecular gas was
concentrated in the central regions, with secondary peaks at the bar
ends due to the bar potential and the viscosity of the gas (\citealt{Regan99,
Sakamoto99}).

Similar to the stellar luminosity profiles, the observed \CO\ radial
distributions in NGC~3627, NGC~3628 and M51 could also be well
fitted by an exponential fit in $R$
 \begin{equation}
I(R) = I_{\rm 0}\ {\rm exp}\ (-R/R_{\rm 0}),
\end{equation}
where $R$ is the galactocentric distance, $R_{\rm 0}$ is the disk
scale length, and $I_{\rm 0}$ is the central integrated intensity.
The solid curves in Fig.~\ref{Fig3} show the least-squares fit to
the data excluding the center point as the nuclear gas could be a
distinct component, yielding $I_{\rm 0}$ = 32.6\,K km s$^{-1}$ and
$R_{\rm 0}$ = 3.5\,kpc for the \CO\ emission in NGC~3628, $I_{\rm 0}$
= 37.8\,K km s$^{-1}$ and $R_{\rm 0}$ = 2.7\,kpc and  $I_{\rm 0}$ =
3.5\,K km s$^{-1}$ and $R_{\rm 0}$ = 3.8\,kpc for \CO\ and $^{13}$CO
emission in M51. In NGC~3627, the scale lengths are 2.8\,kpc and
3.9\,kpc for the \CO\ and $^{13}$CO emission, respectively. However,
the power law fit with functional form, $I = I_{\rm 0}\ R^\alpha$,
is more suitable than exponential fit for the \CO\ distribution
in NGC~4631. The fit result gives $\alpha = -1.5$ and $I_{\rm 0}$ =
41.1\,K km s$^{-1}$. For the other three exponential fit galaxies,
the power law could also fit \CO\ distribution almost equally well. The
exponential \CO\ scale lengths of 2-4\,kpc equal to $\sim$ 0.2 $R_{\rm
25}$, which is consistent with the results in \cite*{Nishiyama01}.
The distributions of \CO\ and $^{13}$CO emission along the
axis with position angle of 90$^{\circ}$ in M82 are similar to the
other four galaxies. The distributions along the major axis are not
shown here, since the observations of M82 were carried out by mapping an
area of $4'\times2.5'$.

\subsubsection{The integrated line intensity ratios}
\label{sss:ratio}

The intensity ratio $\rat$ ranges from 3.3$\pm$0.7 to 12.8$\pm$4.3
for all 36 points both detected in \CO\ and $^{13}$CO emission from
normal spiral galaxies, with mean value of 9.9$\pm$3.0 and
5.6$\pm$1.9 for the central and disk regions, respectively. However,
the average $\rat$ in M82 is about 2.5 times higher than that in the
nucleus of normal spiral galaxies. We use the Equation~(2) in \cite*{Young86}
to calculate the mean value of \rat \ at each
radius. The most prevalent trend is that the \rat \ drops at larger
radius in both the barred and non-barred spiral galaxies that we
have mapped along the major axis in our sample~(Fig. 3). Here we
note that two points~($\sim2.5'$ away from center) are found to have
significantly higher ratios~($\sim$9) in M51. These abnormal
high values in the disks are tended to be observed once the
telescope beams are located in the most active star-forming regions
along spiral arms.

The detection of C$^{18}$O in M82 and M51 also allow us to estimate
the intensity ratio of $^{13}$CO to C$^{18}$O, which ranges between
0.9$\pm$0.7 and 5.3$\pm$2.8 with mean value of 2.9$\pm$1.4. The
ratios measured in the nucleus and disk of M51 are 3.7 and 2.6
respectively, agree well with the results of \cite*{Vila08}, which
claimed first detect C$^{18}$O emission in the center of M51 with
$^{13}$CO/C$^{18}$O ratio of 3.6. Also the similar
$^{13}$CO/C$^{18}$O ratio values have been found in some starburst
galaxies \citep[e.g.,][]{Sage91b,Aalto95}. In addition, our
detection of C$^{18}$O emission in the off-center regions of M51 represents
the first report of such detection for this object by far.

Ongoing observations of dense gas tracers HCO$^+$ and CS, toward the nuclear regions of
NGC~4736, M82, NGC~3628, and M51 by far only yield detection in the
latter three galaxies. Here, we also use these limited dense gas
observations, along with the literature data, to help further analyze the CO isotopic results.
The intensity ratios of $^{13}$CO to HCO$^+$ and $^{13}$CO to CS are
found to show little variations between starburst and normal
galaxies, with average values of 1.3$\pm$0.3 and 3.2$\pm$1.4,
respectively. The observed integrated intensity and line ratios and
the literature data used are listed in Table~\ref{Table3}.

\subsection{Kinematics of CO}
\label{ss:kinematics}

Figure~\ref{Fig4} shows the CO position-velocity ($P-V$) diagrams
along the major axes of the galaxies NGC~3627, NGC~3628, NGC~4631,
and M51, and the $P-V$ diagrams with position angle of 0$^{\circ}$
and 90$^{\circ}$ in M82 as well. It can be seen in Fig.~\ref{Fig4}
that $P-V$ diagrams along the major axes tend to show a gradual
increase of rotation velocity in the inner regions (rigid rotation)
and a nearly constant velocity in the outer regions (differential
rotation).

For NGC~3627 and M51, the $P-V$ spatial velocity maps of $^{13}$CO
are also shown in Fig.~\ref{Fig4}. Obviously, both \CO\ and $^{13}$CO
share essentially similar kinematics and distribution. At each
position where observations were made, the distribution of line
intensity and the velocity range for \CO\ and $^{13}$CO emission are
in good agreement. Accordingly, it could also hint that the line
ratios at each position derived from our observations are relatively
quite reliable.

Figure~\ref{Fig5} shows the CO rotation curve and the variation in
line width along the major axis. Using the mean velocities,
inclination, systemic velocity, and position angle of the major axis
that listed in Table~\ref{Table1}, and by the assumption that the
observed velocity reflect only the systemic motion of the whole
galaxy and circular rotation, the rotation velocity $V_{\rm R}$
could be derived via
 \begin{equation}
V_{\rm R} = (V_{\rm obs} - V_{\rm sys})/{\rm sin}i \ {\rm cos}\theta\, ,
\end{equation}
where $V_{\rm obs}$ is an observed velocity of \CO, $V_{\rm sys}$ is
the systemic velocity of the galaxy, $i$ is the inclination angle,
and $\theta$ is the azimuth measured in the disk plane. The peak
velocity, $V_{\rm max}$, in the rotation curve of our sample
galaxies, ranges from 120 to 240\,km s$^{-1}$ \citep[$V {\rm max}$
in spiral galaxies are usually between 150 and 300\,km
s$^{-1}$,][]{Sparke00}.

\section{Discussion}
\label{sect:discussion}

\subsection{Radial Distribution of Molecular Gas}
\label{ss:radial}

In section~\ref{ss:distribution}, we show that the surface density of
the molecular gas in the galaxies observed in our sample can be well
fitted both by exponential and power law function. The $\chi^2$
values indicate that the data fitting are at about 85\% and 90\%
confidence level by the exponential and power law function, respectively.
We are therefore unable to distinguish clearly between an
exponential and a power law radial distribution over the regions
observed here with limited sampling and low resolution. The similar
conclusion has been pointed out in both \cite*{Young82} and
\cite*{Nishiyama01}. However, \cite*{Scoville83} considered that the
exponential distribution was more suitable than
that of power law for M51. The region in M51 observed by \cite*{Scoville83}
is more extended ($\sim8.5'$) than the region
($\sim7'$) observed here, and thus the exponential functional form
seems to be better for describing the distribution of molecular gas
in the whole galaxy. In NGC~3627 and M51, the scale lengths $R_{\rm
0}$ of the \CO\ profiles agree well with the optical $K$-band scale
lengths of 3.5 and 3.2\,kpc (\citealt{Regan01}), respectively. The
results are in line with the finding first noted by \cite*{Young82}
that the large-scale radial distribution of the \CO\
gas follows the light of the stellar disk. In fact, some recent high
resolution observations also reveal that the \CO\ profiles in a
majority of galaxies follow the two-component (bulge and disk)
stellar brightness distribution quite well (\citealt{Regan01}). The
$^{13}$CO profiles detected in our observations are in good
agreement with that in \cite*{Paglione01}. In NGC~3627 and M51,
$^{13}$CO profiles follow similar exponential distributions as \CO.
Because of insufficient data points with significantly high
signal-to-noise, however, the detected $^{13}$CO data in other
galaxies are too limited to reliably present more useful information
on their distributions.

Comparing the radial distribution of CO integrated intensities in
Fig.~\ref{Fig3} with the $P-V$ map of CO emission intensities in
Fig.~\ref{Fig4}, it is found that the intensities are deficient in
the center of NGC~3627, NGC~4631, and M51. However, the centrally
deficient feature in molecular ring emission is apparent in NGC~3627
and NGC~4631, but not in M51. So combining the variations in line
width shown in Fig.~\ref{Fig5}, we believe the decrement of CO
intensity in the central region of M51 is likely as a result of the
dilution in velocity with much wider line width in the center than
that in the outer regions of the central disk. On the contrary, the
molecular ring emission features with little variations in line
width within the bar region in NGC~3627 and NGC~4631, are probably
either due to orbital disruption at the inner Lindblad resonance or
the central holes with gas exhausted by star formation in the
nuclei. \cite*{Sakamoto99} have modeled the $P-V$ diagrams to
analyze gas distributions, and found that the central gas hole is
easier to find from $P-V$ diagrams than from integrated intensity
maps due to the velocity information.

\subsection{Variations in the Intensity Ratio of \CO\ to $^{13}$CO}
\label{ss:variation}

Previous studies on the variations in \rat \ in external galaxies
have pointed out that some mergers, LIRGs/ULIRGs, and the central
regions of circumnuclear starburst galaxies tend to have higher
values of \rat \ than that in normal spiral galaxies (\citealt{Aalto91,
Aalto95,Casoli92,Greve09,Sage&Isbell91}), of which is similar to that in
giant molecular clouds of our Galaxy. Besides the enhancement of $^{12}$C
caused by nucleosynthesis from type-II
SN of massive stars, the deficiency of $^{13}$CO caused by
isotope-selective photodissociation, and the different distributions
of \CO\ and $^{13}$CO were disputed to be alternative reasons for the
high \rat \ value for a long time (\citealt{Aalto95,Casoli92,Sage91b,
Taniguchi99}). However, the
single-component model calculation of non-LTE CO excitation in
\cite*{Paglione01} suggested that the variations of kinetic
temperature, cloud column and volume density, as well as starburst
superwinds might all contribute to the explanations for the
variation in \rat. We here explore the possible causes of the
observed variations in \rat.

\subsubsection{Possible causes of variations in \rat}
\label{sss:causes}

Our results show that \rat \ varies not only in the nuclei of
various types of galaxies, but also within galaxies between nuclear
regions and outer disks. The Galactic $^{12}$C/$^{13}$C
abundance ratio was found to range from $\sim$30 at 5\,kpc in the
Galactic center to $\sim$70 at 12\,kpc (\citealt{Langer90}).
However, this isotopic gradient is opposite to the gradient in the
molecular abundance ratio \rat \ (Fig. 3). Therefore, the
enhancement of $^{12}$C in starburst regions is unlikely to be an
appropriate explanation for the measured high \rat.

Some authors argue that the selective dissociation of $^{13}$CO
caused by ultraviolet~(UV) field in massive star formation regions
can stimulate the ratio of \rat, since the rarer isotope is less
shielded from photodissociation (\citealt{van88}).
Consequently, C$^{18}$O should be even more dissociated by UV
photons than $^{13}$CO due to its lower abundance, and a
positive correlation between \rat \ and $^{13}$CO/C$^{18}$O
intensity ratio would expect to exist if this is available.
However, Fig.6b shows a very marginal anti-correlation
between \rat\ and $^{13}$CO/C$^{18}$O
with a correlation coefficient $R$=-0.34. Therefore, contrary to the expection,
our results reveal a weak anti-correlation, and the deficiency of $^{13}$CO caused by
isotope-selective photodissociation could be ruled out.

In addition to the high $\rat$ measured in M82, the integrated
$J$=2-1/$J$=1-0 line ratio was found to range between 1.0 and 1.4
(\citealt{Mao00}), revealing the existence of highly excited
molecular gas. In the PDR model of \cite*{Mao00}, it was suggested
that the bulk of \CO\ emission arises from warm diffuse interclumpy
medium whereas $^{13}$CO emission originate in denser cores. In the
warm PDRs, the optical depth of \CO\ emission from the envelope gas
could decrease to a moderate value of \ $\tau\sim$1, result in the
corresponding $\tau(^{13}{\rm CO})<<$ 1. Moreover, the large
concentrations of molecular gas in the nuclear starburst with high
$T_{\rm K}$ can be excited by strong UV emission from young massive
stars, shocks and turbulence caused by supernova remnants, and
cosmic rays (\citealt{Pineda08,Zhang10}). The most recent $Herschel$
observations of M82 also suggested that turbulence from stellar
winds and supernova may dominate the heating of warm molecular gas
in the central region (\citealt{Panuzzo10}). Therefore, it is likely
to imply that nonthermal motions produced by the stellar superwinds
can broaden \CO\ lines, thus enhance \CO\ emission as more photons
located deeper in the clouds are allowed to escape. Furthermore, the
significant high value of \rat \ observed in the spiral arms of M51
also demonstrate that \CO\ emission can be enhanced in active
star-forming regions compared with that in inter-arm regions.

\subsubsection{$X$-factor and dense gas ratio in extreme starburst}
\label{sss:extreme}

Both theoretical and observational investigations have revealed that
\CO\ emission is influenced by the intrinsic properties of molecular
cloud \citep[e.g.,][]{Pineda08,Goodman09,Shetty11}. In the
magneto-hydrodynamic models of \cite*{Shetty11}, \CO\ integrated
intensity is found to be not an accurate measure of the amount of
molecular gas, even for $^{13}$CO, which may also not be an ideal
tracer in some regions. In this case, the much lower opacity
C$^{18}$O can give much more reliable constraints on H$_2$ column
density than optically thick \CO\ isotopes. Comparing the H$_2$
column density derived from \CO\ with that from C$^{18}$O listed in
Table~\ref{Table2} and Table~\ref{Table3}, we find that the amount
of molecular gas estimated by standard Galactic $X$-factor is
consistent with that derived from C$^{18}$O in M51, whereas is
overestimated in M82 by a factor of 2.5, equal to the ratio of \rat
\ between M82 and M51. Consequently, our results confirm that the
$X$-factor adopted in starburst active regions should be lower than
that in normal star-forming regions, and the gradient in \rat \ can
trace the variations in $X$-factor. However, surveys of C$^{18}$O in
a larger sample are required to confirm the relation between the
variations in \rat \ and $X$-factor found in this study.

The average ratio of $^{13}$CO to C$^{18}$O ($\sim$2.9$\pm$1.4)
derived from our observations in M51 and M82 indicates that a
portion of $^{13}$CO emission has a moderate optical depth, since
the ratio of $^{13}$CO to C$^{18}$O should be $\sim$7 if both
$^{13}$CO and C$^{18}$O lines are optically thin (\citealt{Penzias81}).
This result is in line with the two-type cloud model suggested in
\cite*{Aalto95}, in which a large fraction of $^{13}$CO
emission might originate from denser gas component. Some previous surveys
in dense molecule have provided support for the presence of such dense
gas \citep[e.g.,][]{Sage90,Nguyen92,Gao04a}. In addition, our detection
of HCO$^+$ and CS in M82 and M51 is consistent with that
in \cite*{Sage90} and \cite*{Naylor10}. The ratios of $^{13}$CO to HCO$^+$ and $^{13}$CO
to CS in (U)LIRGs NGC~3256, NGC~6240 and Arp~220 (\citealt{Casoli92,Greve09})
are found to be lower than those in the
nuclear regions of normal spirals and M82 (Fig.~\ref{Fig6}), which
probably indicate that the dense gas fraction is higher for these
(U)LIRGs, since $^{13}$CO can be considered as a relative reliable
tracer of total molecular gas due to its low abundance. This result agrees
well with the conclusion in \cite*{Gao04b} that galaxies
with high star formation efficiency tend to have higher dense gas
fraction. Moreover, the ratio \rat \ measured in NGC~3256, NGC~6240
and Arp~220 are found to be much higher than those in
M82~(Fig.~\ref{Fig6}), which is likely to imply that the bulk of \CO\
emission arise from warm diffuse gas is enhanced in the extreme
starburst.

Summarizing the above analysis, the systematical gradient in \rat \
can be explained by the variations in the physical conditions of molecular
gas. The standard Galactic $X$-factor used in M82 overestimates the amount of
molecular gas by a factor of 2.5, and the variations in $X$-factor
can be well traced by the gradient in \rat. Nevertheless,
additional observations of both \CO\ and $^{13}$CO lines at J $\geq$ 2
and more C$^{18}$O lines are required to better constrain the
physical conditions of the molecular gas in external galaxies.

\section{Summary}
\label{sect:summary}

We observed simultaneously the \CO, $^{13}$CO and C$^{18}$O emission
lines in 11 nearby infrared-brightest galaxies, of which four
(NGC~3627, NGC~3628, NGC~4631 and M51) were mapped with half-beam
spacing along the major axes and M82 was fully mapped in an area of
$4'\times 2.5'$. These are the first systematic extragalactic
observations for the PMO 14m telescope and the main results are
summarized as follows:

1.We detected the \CO\ emission towards 99 of the positions observed,
with the $^{13}$CO seen towards 51 of these. C$^{18}$O was
detected at 13 positions close to the nuclear regions in M51 and M82,
among which the off-center positions in M51 were the first C$^{18}$O
detection reported here.

2.In the four galaxies with major axes mapping, the $^{13}$CO line
intensity decrease from center to outer disk, similar to that of \CO.
In NGC~3627, NGC~3628, and M51, the radial distribution of both \CO\
and $^{13}$CO can be well fitted by an exponential function, whereas
the \CO\ distribution in NGC~4631 is better fitted by power law. The
scale length of \CO\ emission is about 0.2 $R_{25}$ with a mean value
of $\sim$3\,kpc. Moreover, the \CO\ scale lengths in NGC~3627 and M51
are in good agreement with the optical scale lengths.

3.The peak velocity of \CO\ rotation curves ranges from 120 to 240\,km
s$^{-1}$, and the line widths of \CO\ lines tend to drop with radius
from center to outer disk in all mapped galaxies. Of all positions
observed, the distribution of both line intensity and profiles of \CO\
and $^{13}$CO have good agreement, as expected with simultaneously
\CO\ and $^{13}$CO observations. Thus, a reliable line intensity
ratio \rat \ can be obtained.

4.A decreasing tendency of \rat \ with radius from center to outer
disk is found in mapped galaxies. \rat \ varies from 3.3$\pm$0.7 to
24.8$\pm$2.5 in positions with both \CO\ and $^{13}$CO detected. The
average \rat \ are 9.9$\pm$3.0 and 5.6$\pm$1.9 in the center and
disk regions of normal spiral galaxies,respectively.

5.The high \rat \ measured in M82 is likely to be caused by
enhanced \CO\ emission from deeper cloud with broad \CO\ line produced
by the stellar winds and supernova. The low values of $^{13}$CO/C$^{18}$O
($\sim$2.8$\pm$1.2) found in M82 support the suggestion that a
considerable fraction of $^{13}$CO emission originates in denser
gas component. Comparing the ratios of $^{13}$CO/HCO$^+$ and $^{13}$CO/CS in
normal galaxies with those in U/LIRGs, the lower values found in
U/LIRGs agree with the notion that the galaxies with high SFEs tend
to have higher dense gas fraction.

6.Comparing with the H$_2$ column density derived from C$^{18}$O,
the standard Galactic $X$-factor is found to overestimate the amount
of molecular gas in M82 by a factor of $\sim$2.5. This confirms the
assertion that a lower $X$-factor should be adopted in active
starburst regions than that in normal star-forming disks, and
moreover, the gradient in \rat \ can be used reliably to trace the
variations of the $X$-factor.

\normalem
\begin{acknowledgements}
We thank the staff of Qinghai station for their continuous help. We
are grateful to Thomas Greve and anonymous referee for helpful comments.
This work was funded by NSF of China (Distinguished Young Scholars \#10425313,
grants \#10833006 and \#10621303) and Chinese Academy of Sciences' Hundred
Talent Program.
\end{acknowledgements}


\nocite{*}
\bibliographystyle{spr-mp-nameyear-cnd}
\bibliography{biblio-u1}

\clearpage

\begin{figure}[htb]
\centering
\makeatletter
\renewcommand{\thefigure}{\@arabic\c@figure{}a}
\makeatother
\begin{minipage}[c]{\textwidth}
\centering
\includegraphics[scale=0.4]{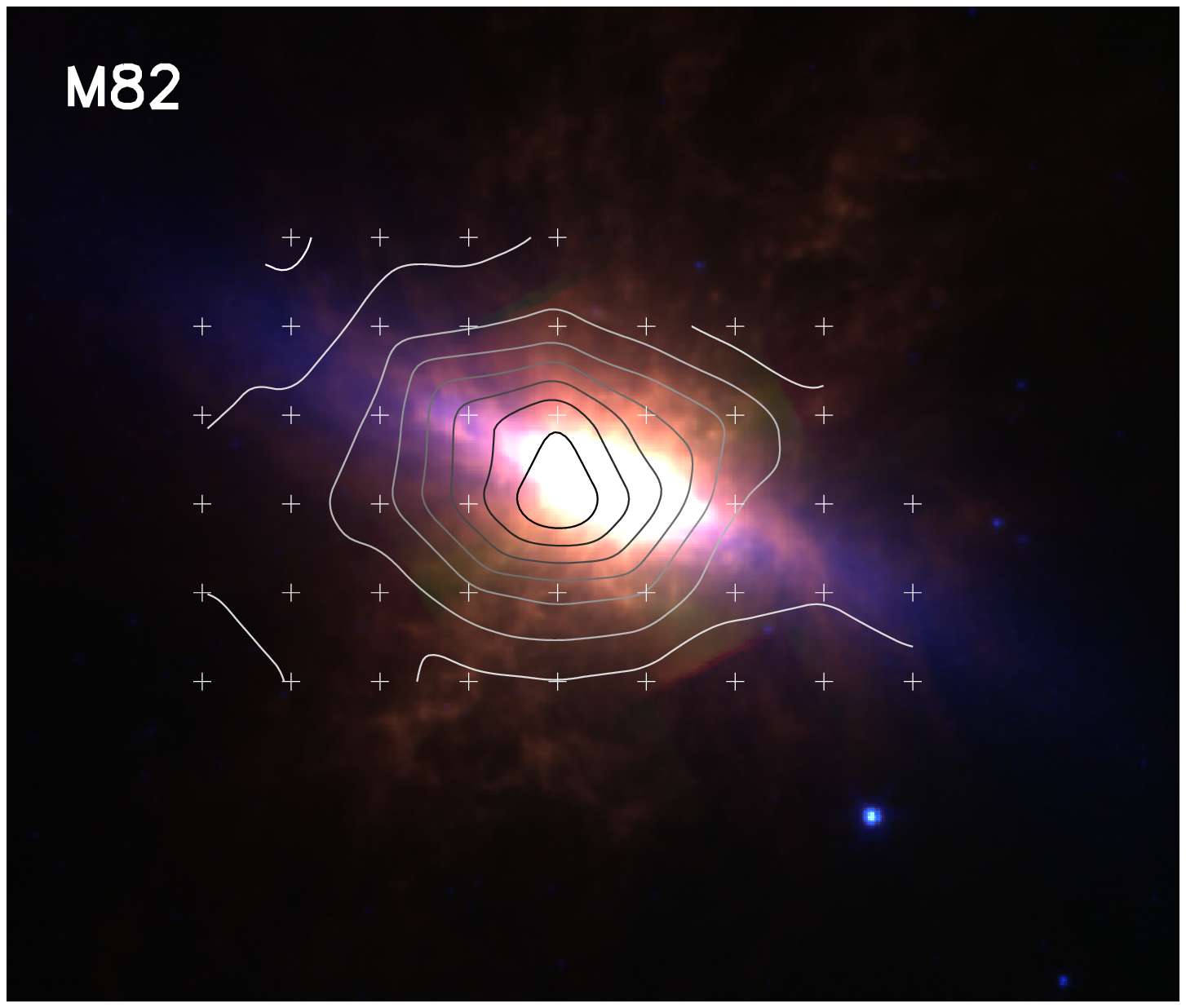}
\end{minipage}\\[10pt]
\begin{minipage}[t]{\textwidth}
\centering
\includegraphics[angle=-90,width=\textwidth]{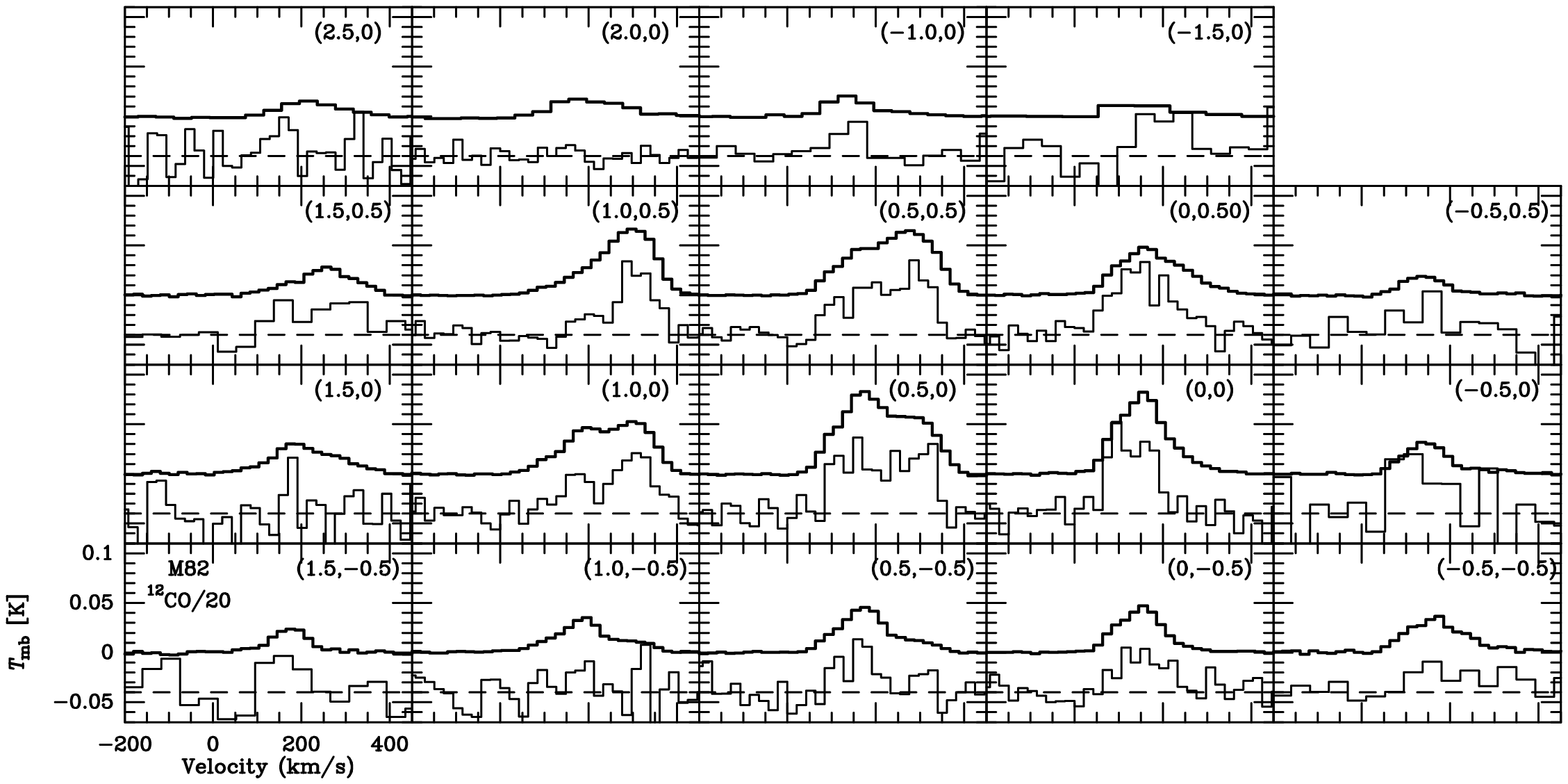}
\end{minipage}
\caption{Spectra of \CO\ ($thick \ lines$) and $^{13}$CO ($thin \
lines$) from the central region of M82 with $^{13}$CO tentatively
detected. All spectra are on the $T_{\rm mb}$ scale and binned to a
velocity resolution of $\sim$20\,km s$^{-1}$ (for some weak
$^{13}$CO emission positions, the spectra are further smoothed to
$\sim$40\,km s$^{-1}$ for display). \CO\ spectra are divided by 20
for comparison purposes. The offset from the center position is
indicated in each box. A linear baseline has been subtracted using
the line-free portions of each spectrum. M82 was mapped in the
$4'\times 2.5'$ central region. The top panel shows the observed
positions(crosses) and \CO\ contours (contours begin at 10\ K km
s$^{-1}$ and increase by 40\ K km s$^{-1}$ each step) overlaid on
infrared image taken from {\sl Spitzer} (8.0\ $\mu$m ($red$), 5.8\
$\mu$m ($green$), 3.6\ $\mu$m ($blue$)) }\label{Fig1}
\end{figure}

\begin{figure}[htb]
\centering
\addtocounter{figure}{-1}
\makeatletter
\renewcommand{\thefigure}{\@arabic\c@figure{}b}
\makeatother
\begin{minipage}[t]{0.5\textwidth}
\vspace{0pt}
\centering
\includegraphics[width=0.7\textwidth]{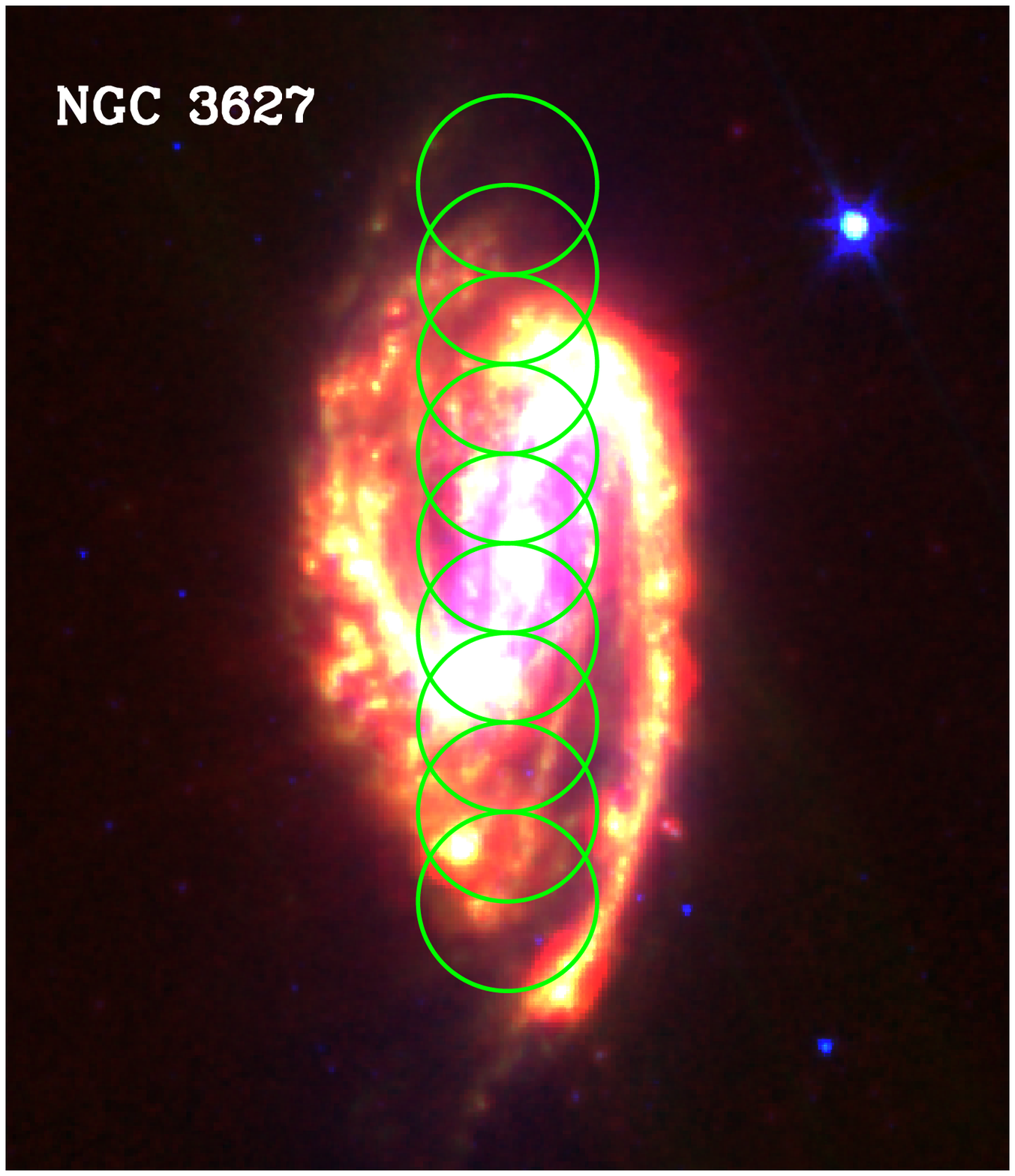}
\end{minipage}%
\begin{minipage}[t]{0.5\textwidth}
\vspace{0pt}
\centering
\includegraphics[angle=-90,width=\textwidth]{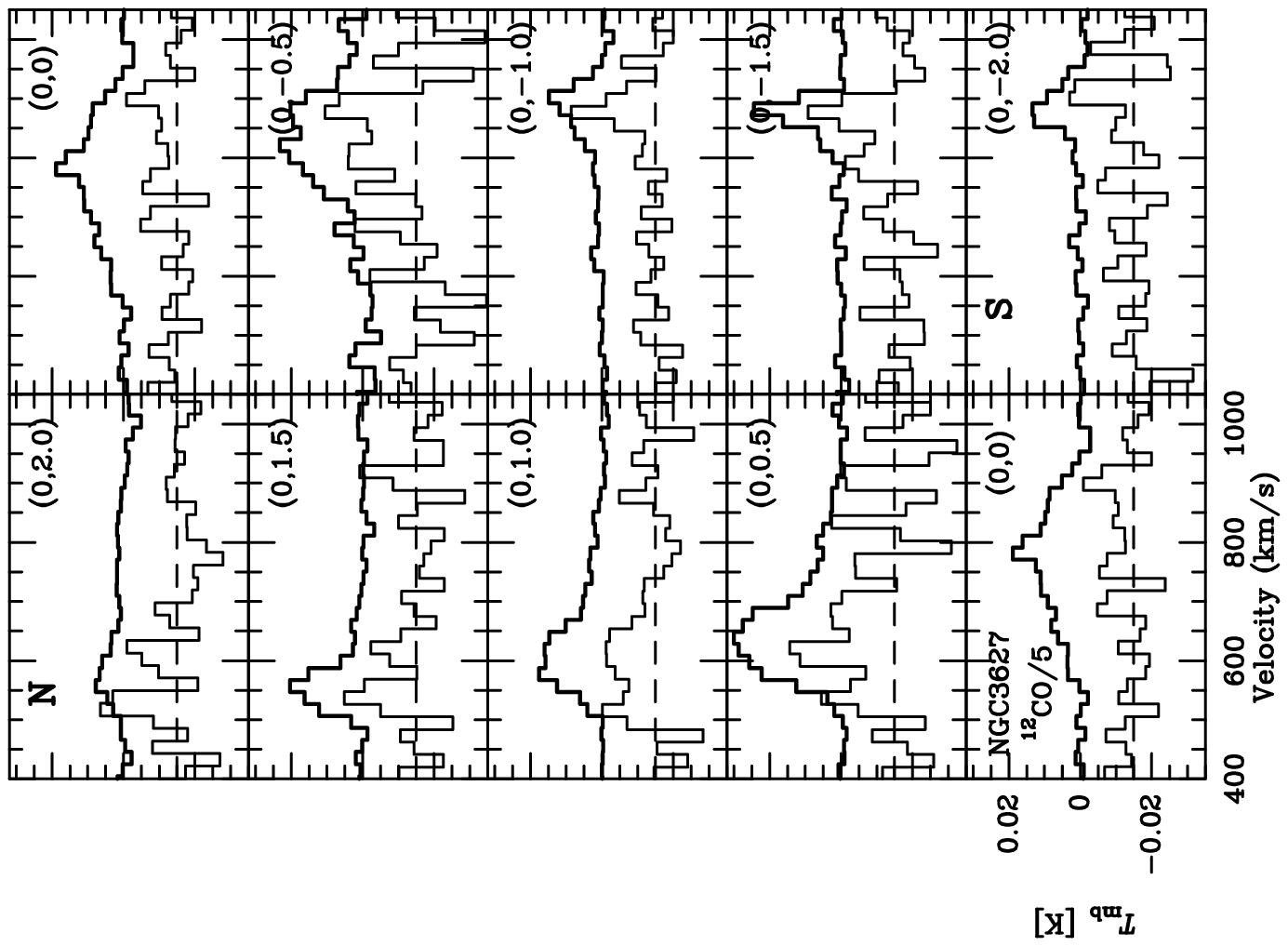}
\end{minipage}
\caption{Same as figure 1a, but for NGC 3627. The spectra were measured
along the major ($PA$=176$^\circ$) axes of the galactic disk. The left panel
shows the half-beam spacing observations overlaid on {\sl Spitzer} infrared
image~(24.0$\mu$m ($red$),8.0$\mu$m ($green$),3.6$\mu$m ($blue$)). The size
of HPBW ($\sim$60$''$) is represented by circle.}\label{fig1b}
\end{figure}

\begin{figure}[htb]
\centering
\addtocounter{figure}{-1}
\makeatletter
\renewcommand{\thefigure}{\@arabic\c@figure{}c}
\makeatother
\begin{minipage}[t]{0.5\textwidth}
\vspace{0pt}
\centering
\includegraphics[width=0.85\textwidth]{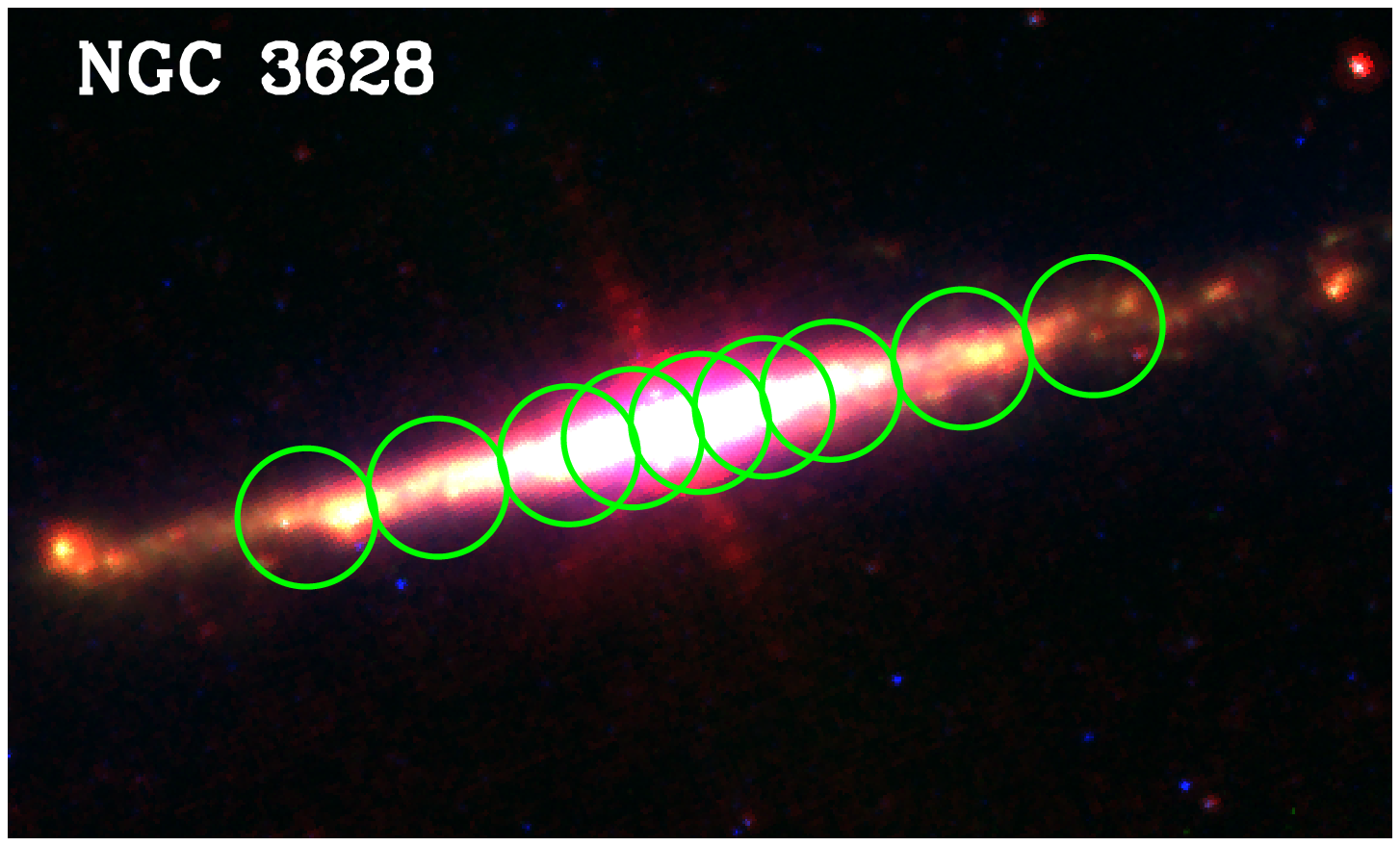}
\end{minipage}%
\begin{minipage}[t]{0.5\textwidth}
\vspace{0pt}
\centering
\includegraphics[angle=-90,width=\textwidth]{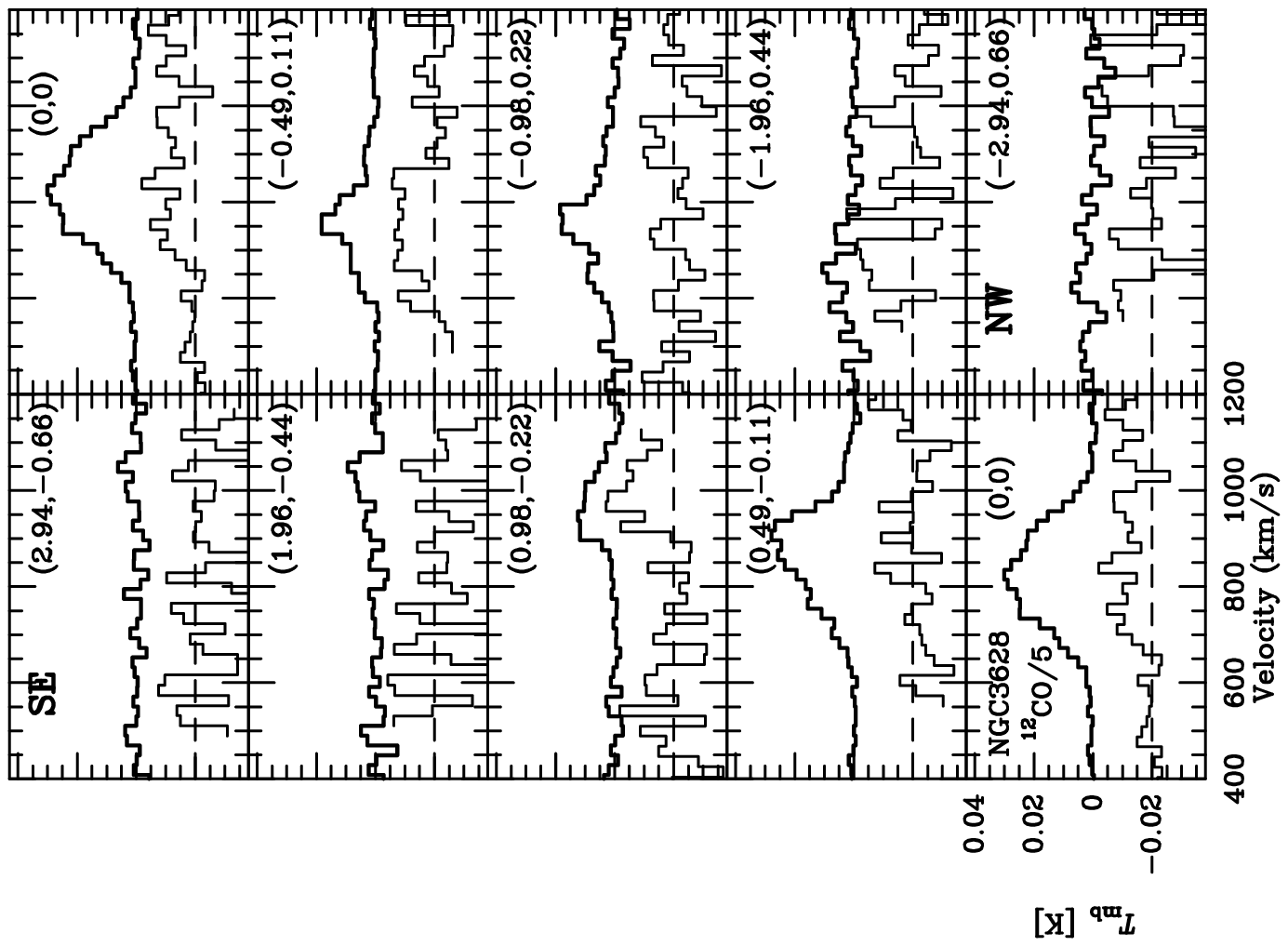}
\end{minipage}
\caption{Same as figure 1b, but for NGC 3628. The spectra were measured along the
major ($PA$=103$^\circ$) axes of the galactic disk.}\label{fig1c}
\end{figure}

\begin{figure}[htb]
\centering
\addtocounter{figure}{-1}
\makeatletter
\renewcommand{\thefigure}{\@arabic\c@figure{}d}
\makeatother
\begin{minipage}[t]{0.4\textwidth}
\vspace{0pt}
\centering
\includegraphics[width=\textwidth]{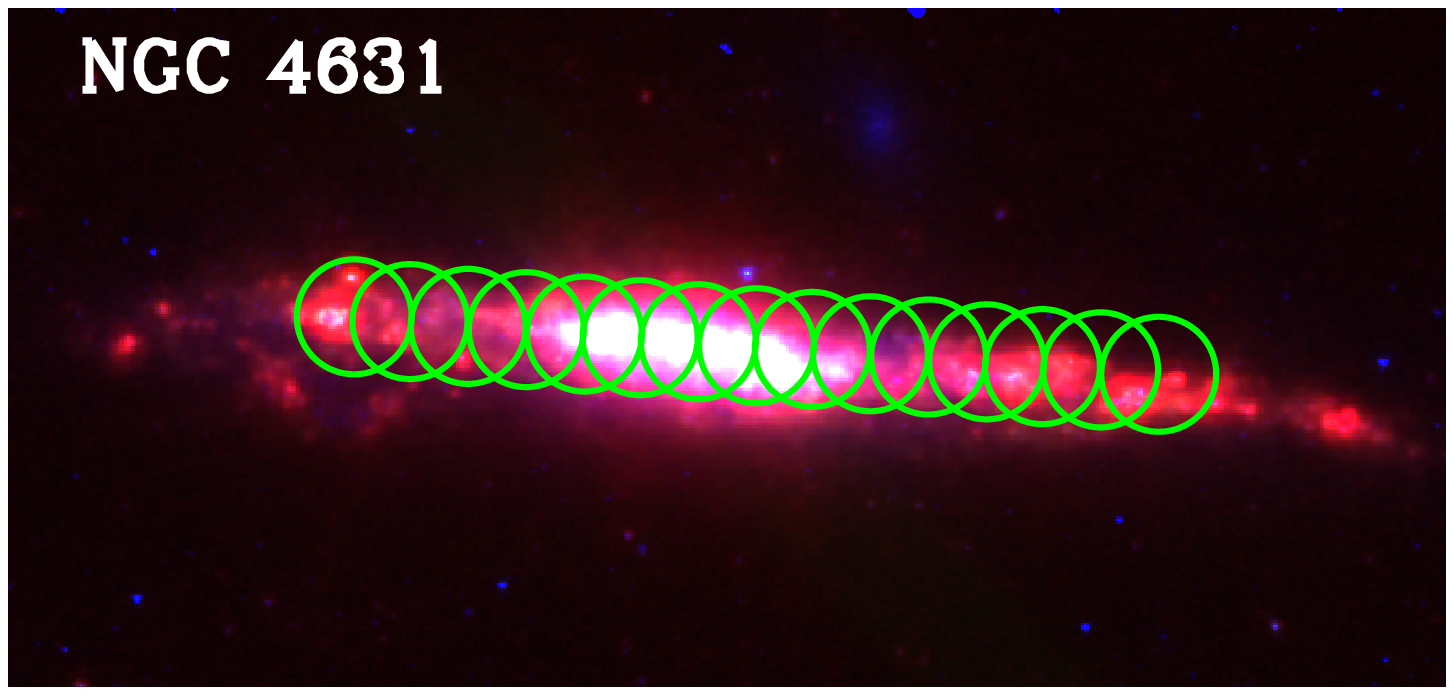}
\end{minipage}%
\begin{minipage}[t]{0.6\textwidth}
\vspace{0pt}
\centering
\includegraphics[angle=-90,width=\textwidth]{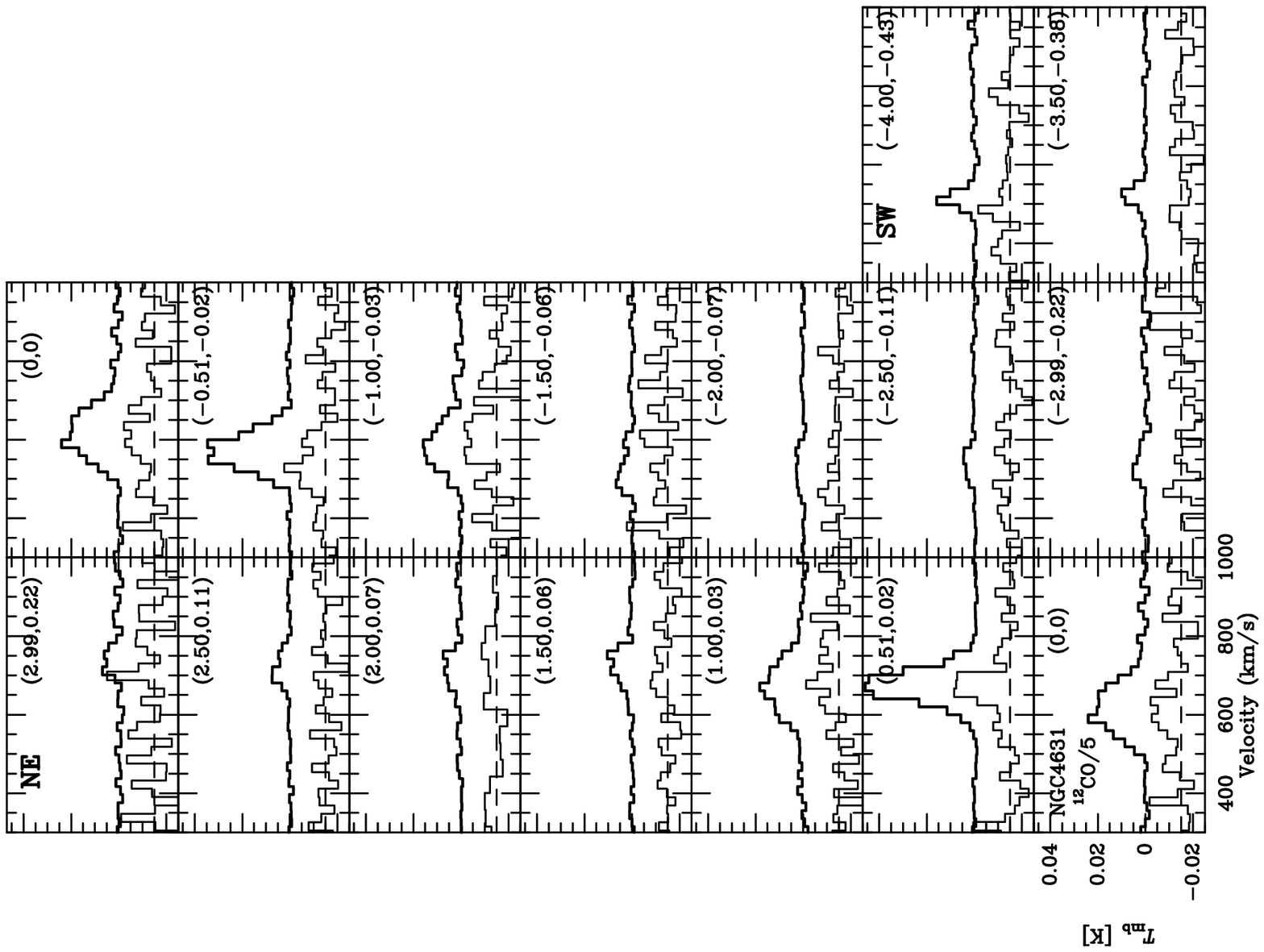}
\end{minipage}
\caption{Same as figure 1b, but for NGC 4631. The spectra were measured along the
major ($PA$=88$^\circ$) axes of the galactic disk.}\label{fig1d}
\end{figure}

\begin{figure}[htb]
\centering
\addtocounter{figure}{-1}
\makeatletter
\renewcommand{\thefigure}{\@arabic\c@figure{}e}
\makeatother
\begin{minipage}[t]{0.5\textwidth}
\vspace{0pt}
\centering
\includegraphics[width=0.8\textwidth]{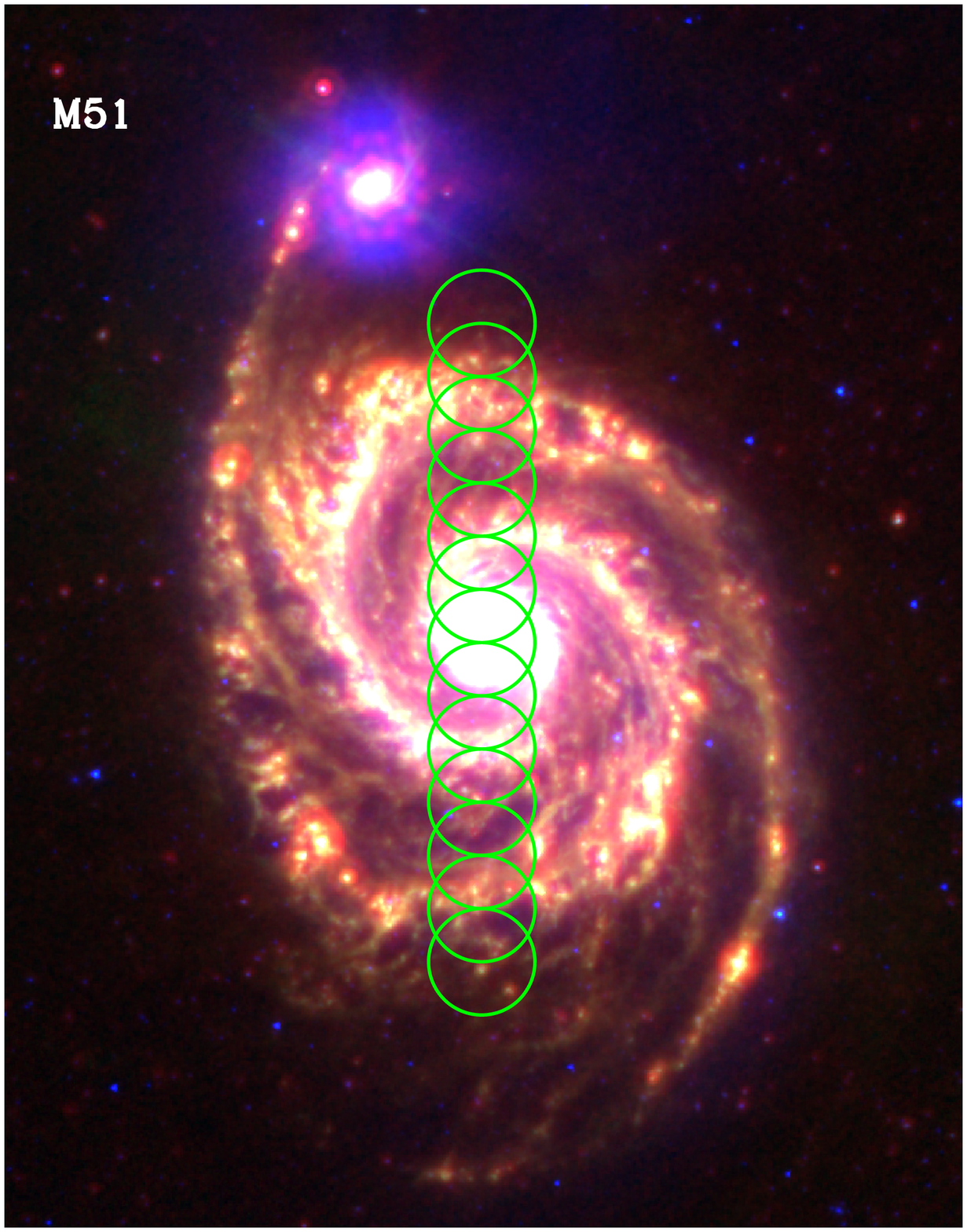}
\end{minipage}%
\begin{minipage}[t]{0.5\textwidth}
\vspace{0pt}
\centering
\includegraphics[angle=-90,width=\textwidth]{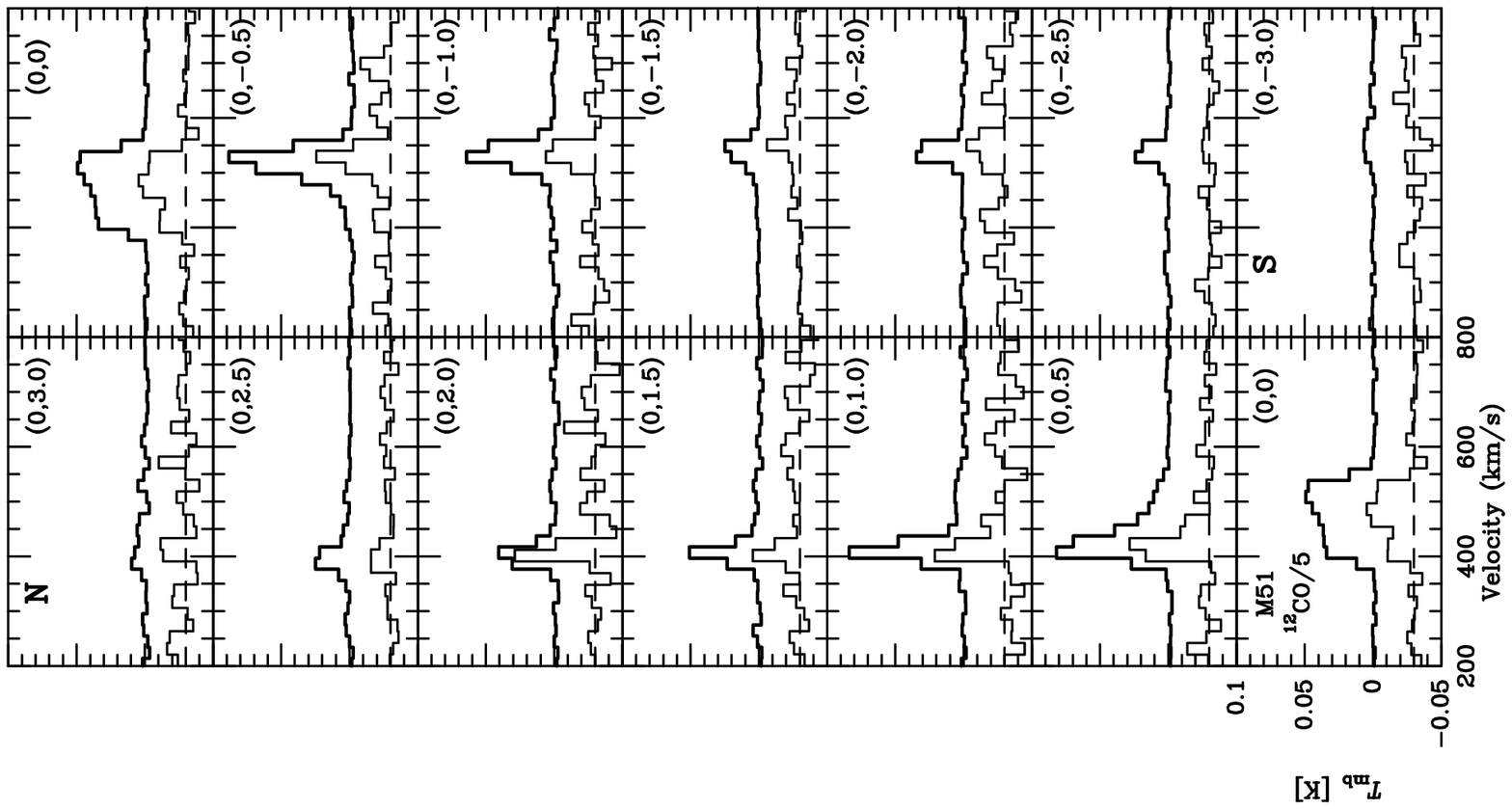}
\end{minipage}
\caption{Same as figure 1b, but for m51. The spectra were measured along the
major ($PA$=0$^\circ$) axes of the galactic disk.}\label{fig1e}
\end{figure}


\begin{figure}[htbp]
\begin{minipage}[t]{\textwidth}
\includegraphics[angle=-90,width=\textwidth]{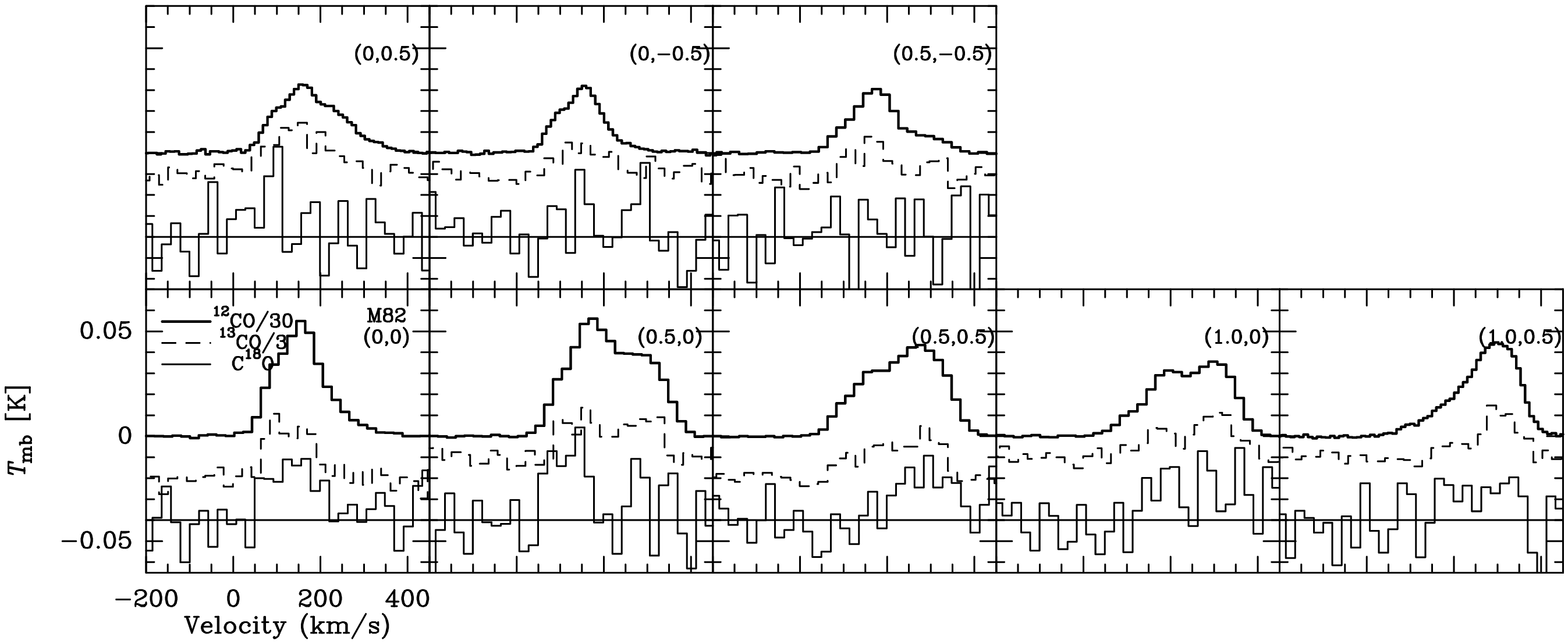}
\end{minipage}\\[4pt]
\begin{minipage}[t]{\textwidth}
\includegraphics[angle=-90,width=\textwidth]{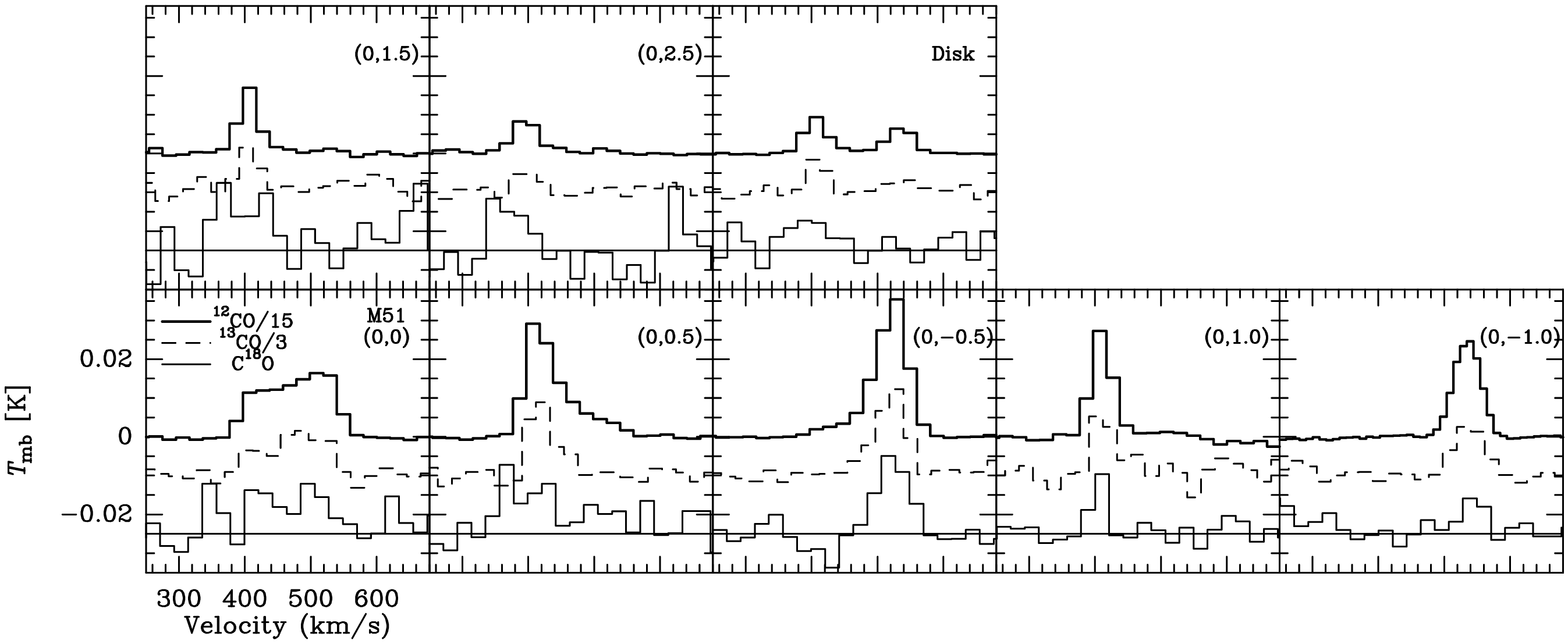}
\end{minipage}

\caption{Spectra of $^{12}$CO, $^{13}$CO and C$^{18}$O obtained from
the same positions on the galaxy M82 and M51. $^{12}$CO spectra are
divided by 30 and 15 for display in M82 and M51, while $^{13}$CO
spectra are divided by 3 for display. M51 disk spectra represent the
average emission over the disk region except the center (0,0).}\label{Fig2}
\end{figure}


\begin{figure}[htbp]
\begin{minipage}[t]{0.33\textwidth}
\centering
\includegraphics[scale=0.7]{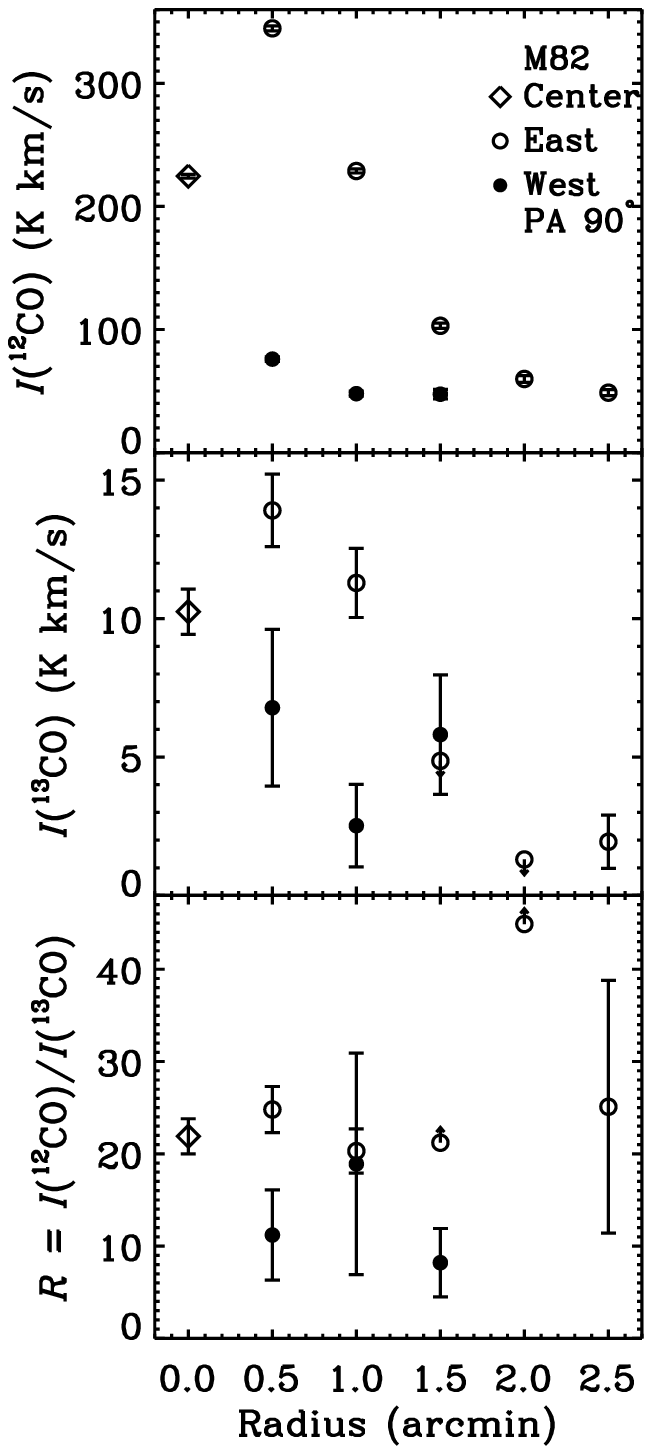}
\end{minipage}%
\begin{minipage}[t]{0.33\textwidth}
\centering
\includegraphics[scale=0.7]{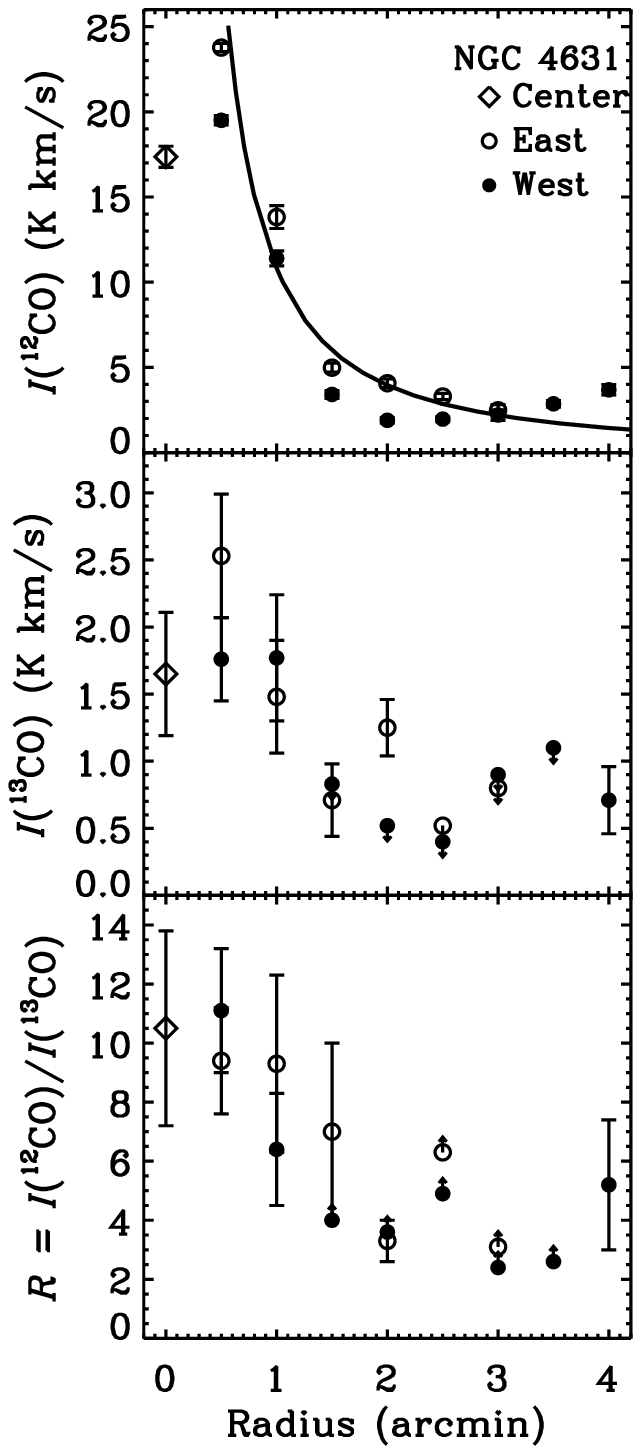}
\end{minipage}\\[10pt]
\begin{minipage}[t]{0.33\textwidth}
\centering
\includegraphics[scale=0.7]{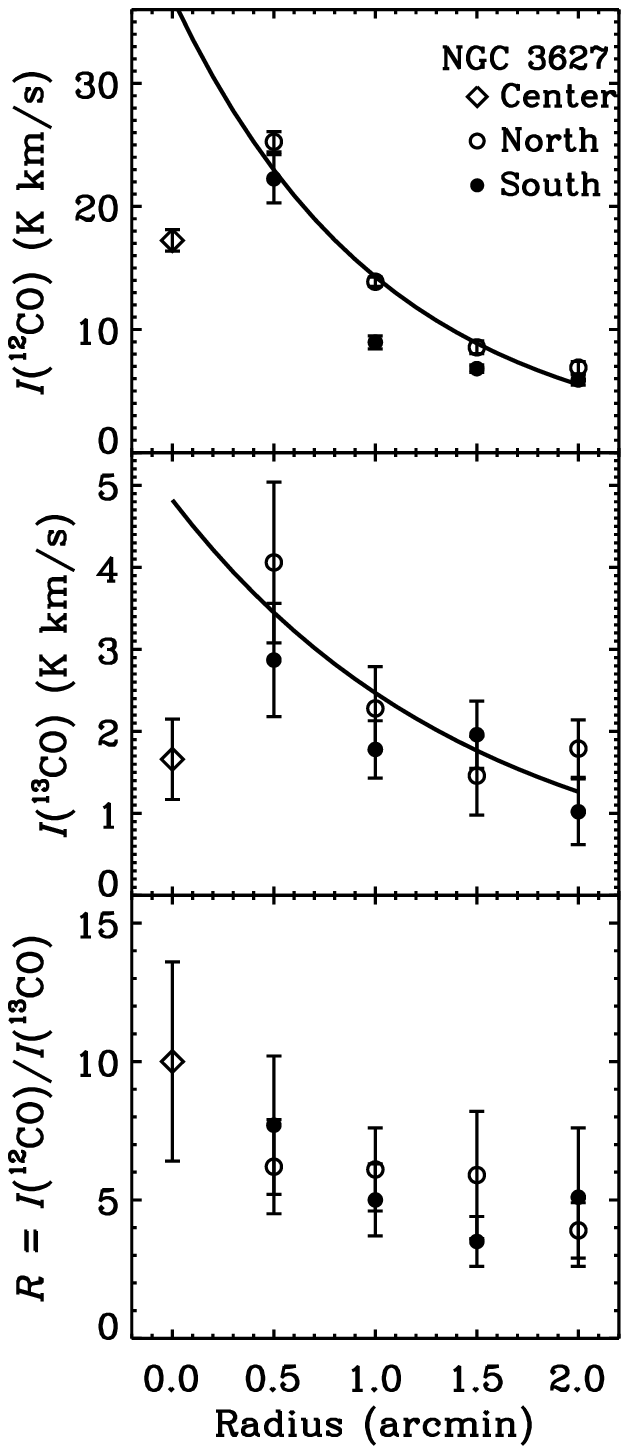}
\end{minipage}%
\begin{minipage}[t]{0.33\textwidth}
\centering
\includegraphics[scale=0.7]{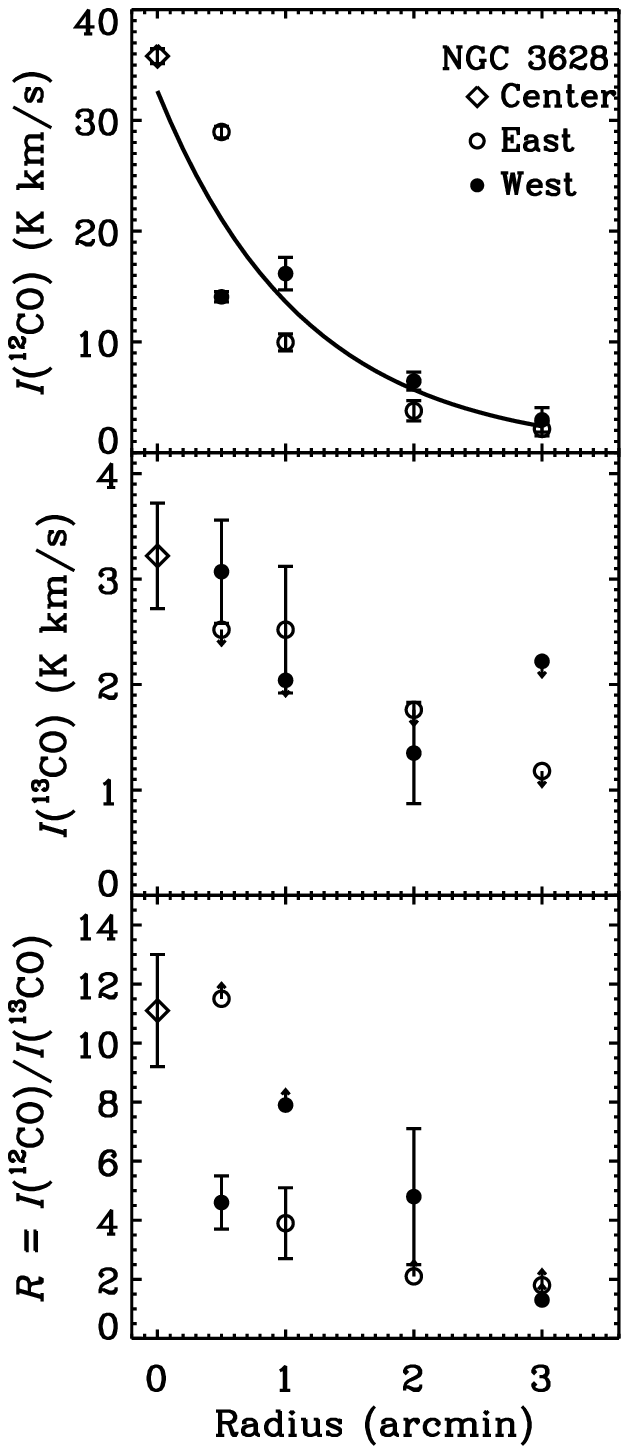}
\end{minipage}%
\begin{minipage}[t]{0.33\textwidth}
\centering
\includegraphics[scale=0.7]{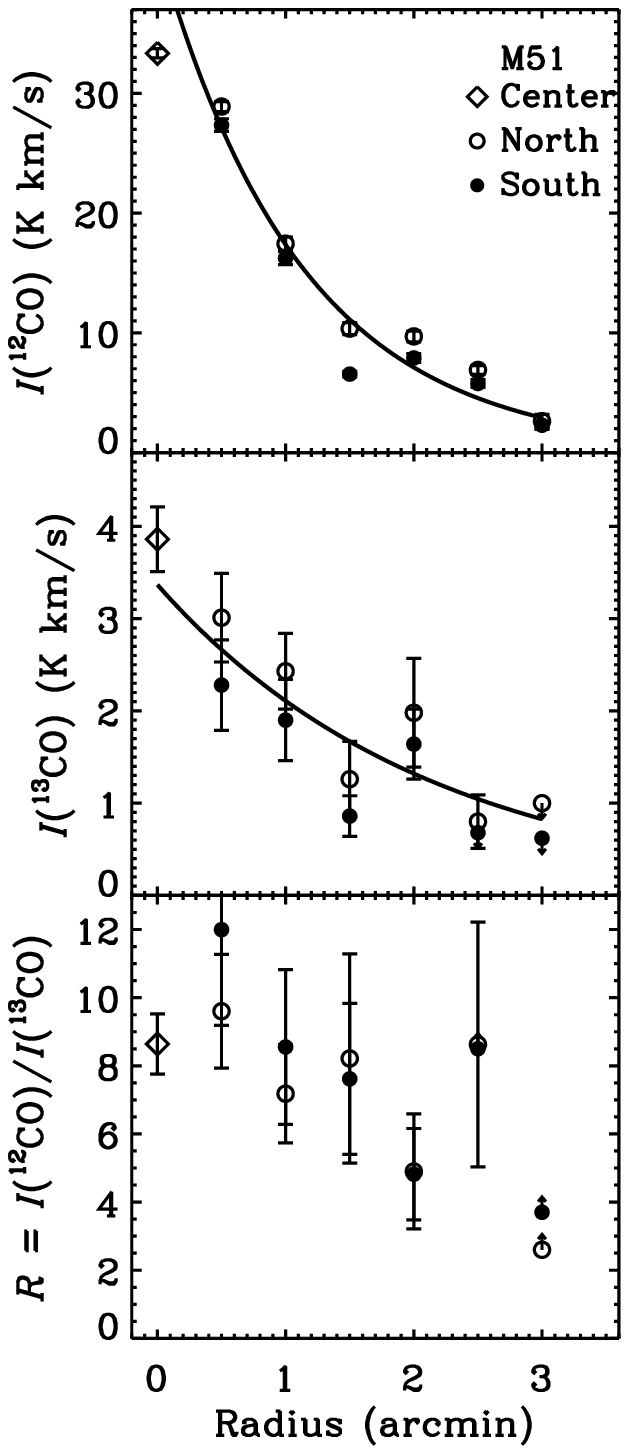}
\end{minipage}

\caption{Radial distributions of $^{12}$CO and $^{13}$CO integrated
intensity and their ratios at each position along the major axes
(for M82,the distributions along the axis with position angle of 90
degree are shown). Error bars are 1 $\sigma$ statistical uncertainty
based on the measured rms noise in each spectrum. Upper limits
(2$\sigma$) are denoted with downward arrow for the non-detection
$^{13}$CO emission, the corresponding lower limits of line ratio
$\rat$ are denoted with upward arrow. The solid line in NGC~3627,
NGC~3628 and M51 represent the exponential fit to the radial
distribution of mean intensity of \CO\ and $^{13}$CO emission, while
in NGC~4631 represents the power law fit to \CO\ emission.}\label{Fig3}
\end{figure}

\begin{figure}[htbp]
\begin{minipage}[t]{0.33\textwidth}
\includegraphics[angle=-90,width=\textwidth]{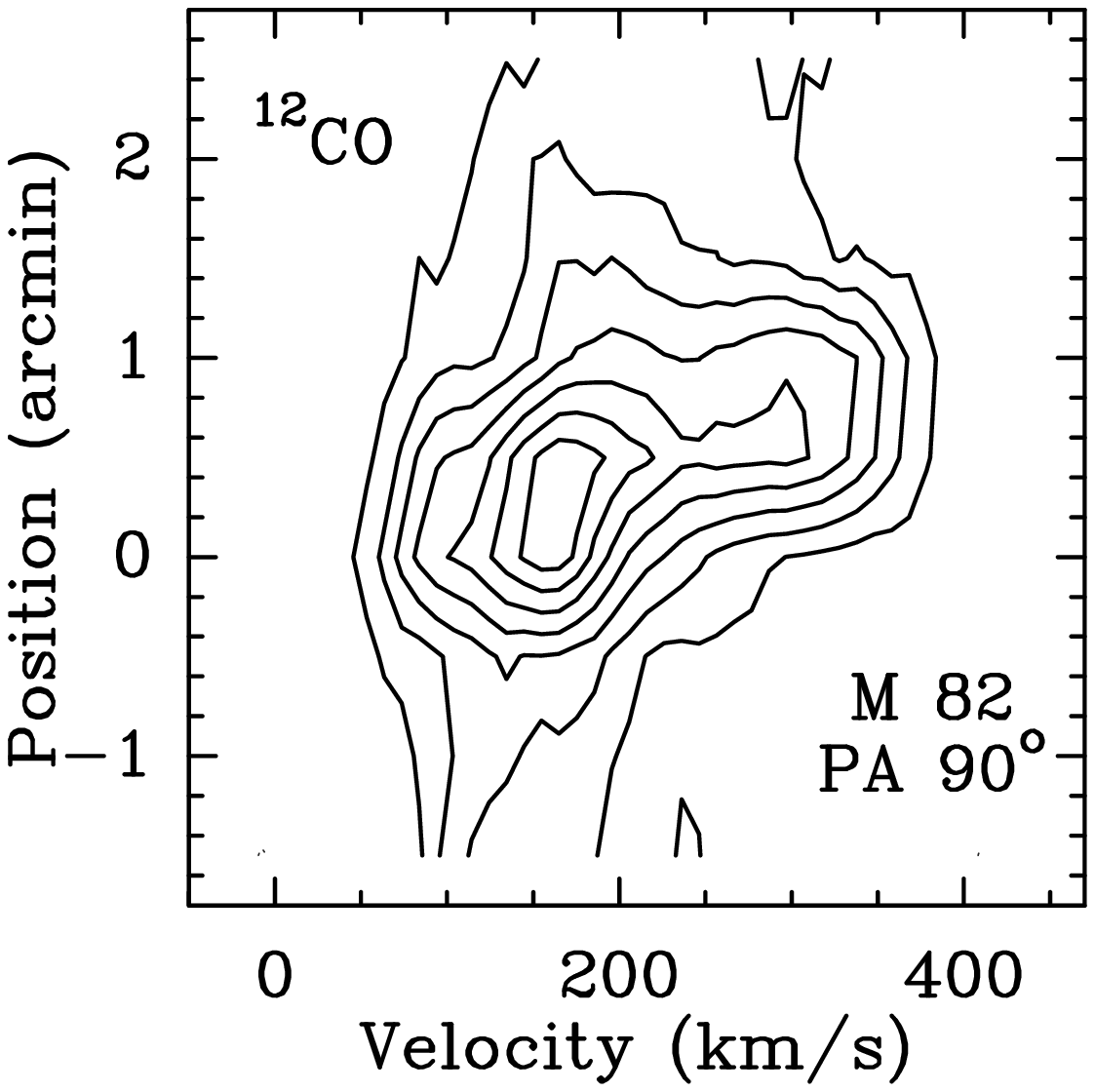}
\end{minipage}%
\begin{minipage}[t]{0.33\textwidth}
\includegraphics[angle=-90,width=\textwidth]{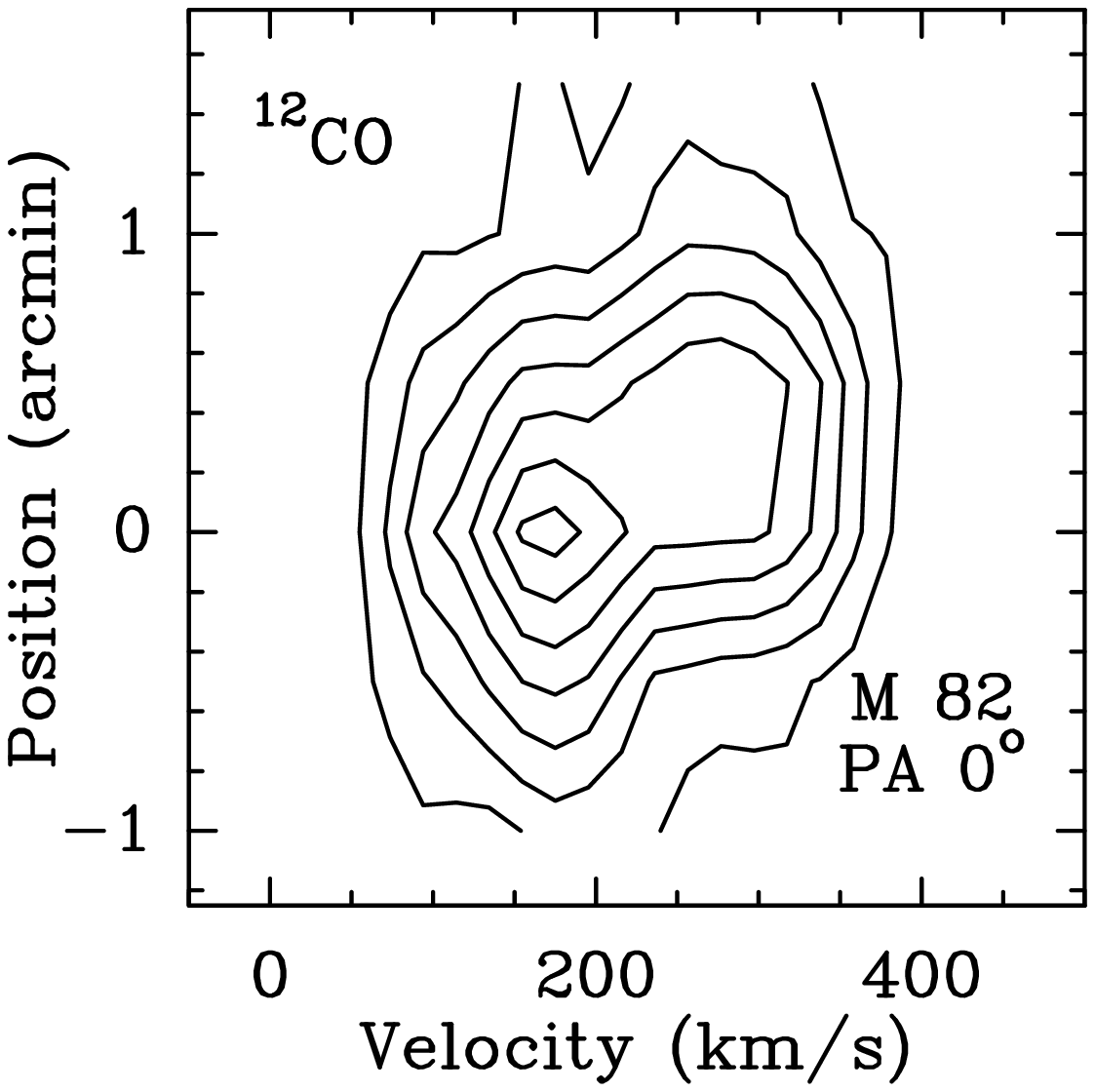}
\end{minipage}\\[10pt]
\begin{minipage}[t]{0.66\textwidth}
\includegraphics[angle=-90,width=0.91\textwidth]{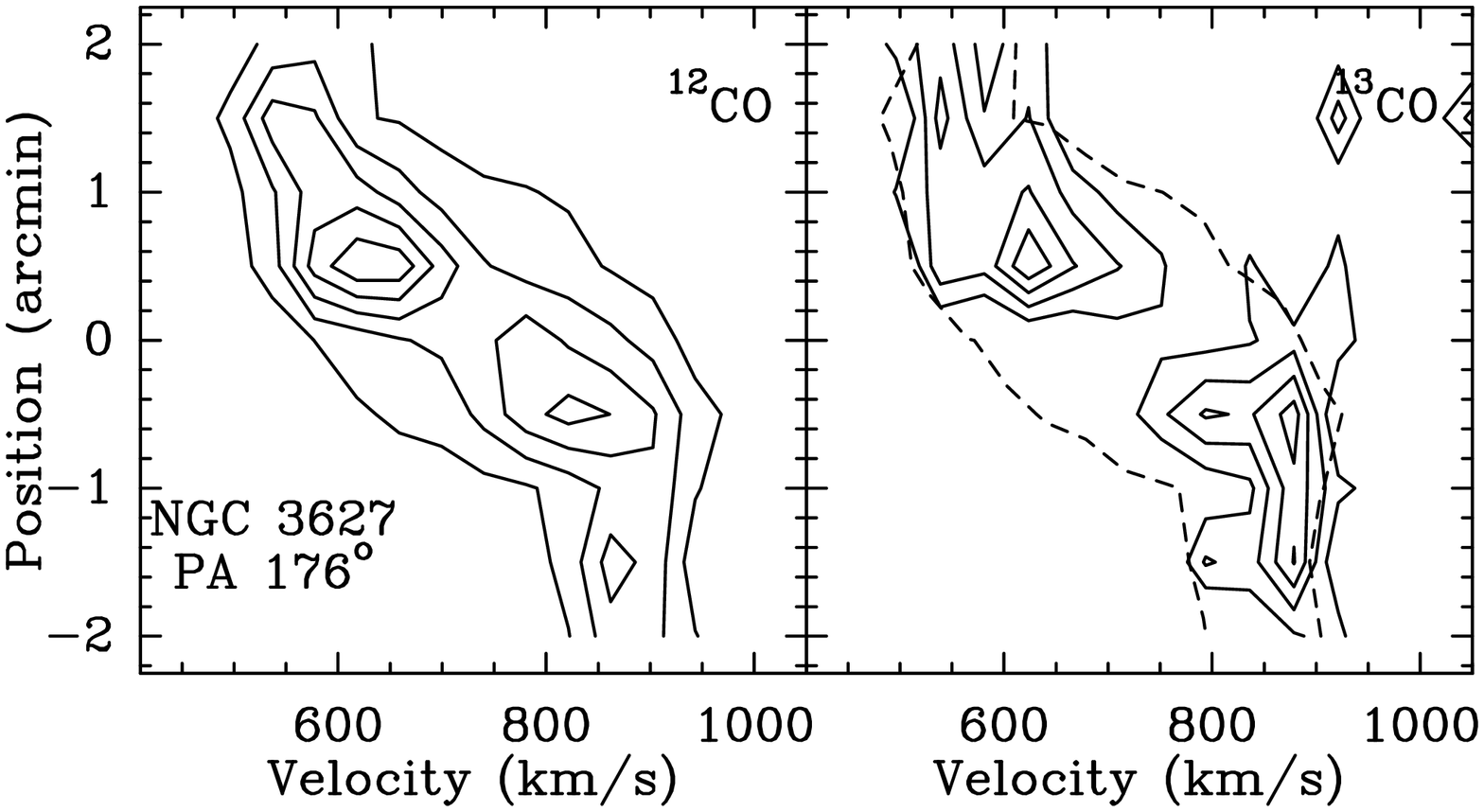}
\end{minipage}%
\begin{minipage}[t]{0.33\textwidth}
\includegraphics[angle=-90,width=\textwidth]{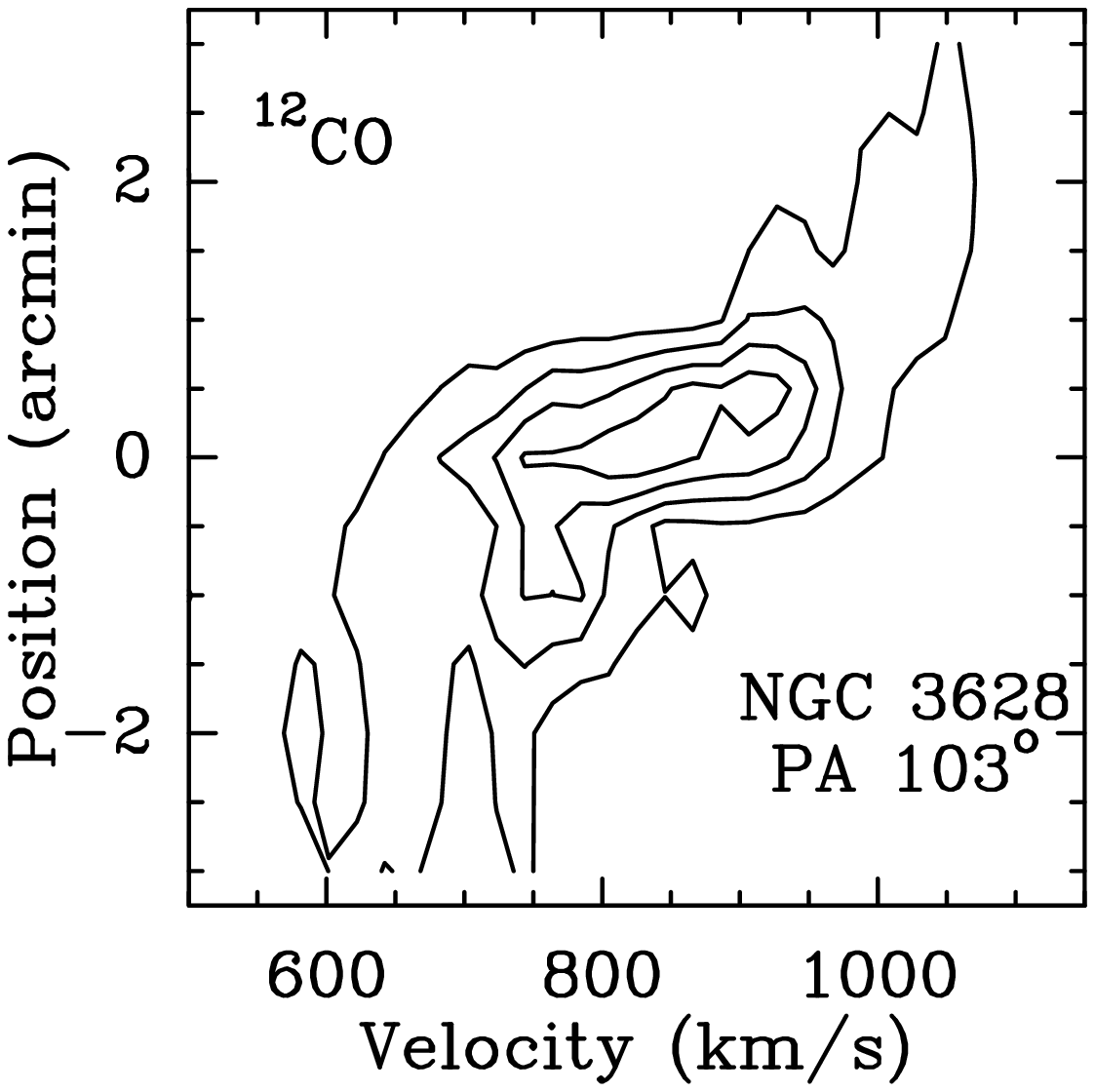}
\end{minipage}\\[10pt]
\begin{minipage}[t]{0.66\textwidth}
\includegraphics[angle=-90,width=0.91\textwidth]{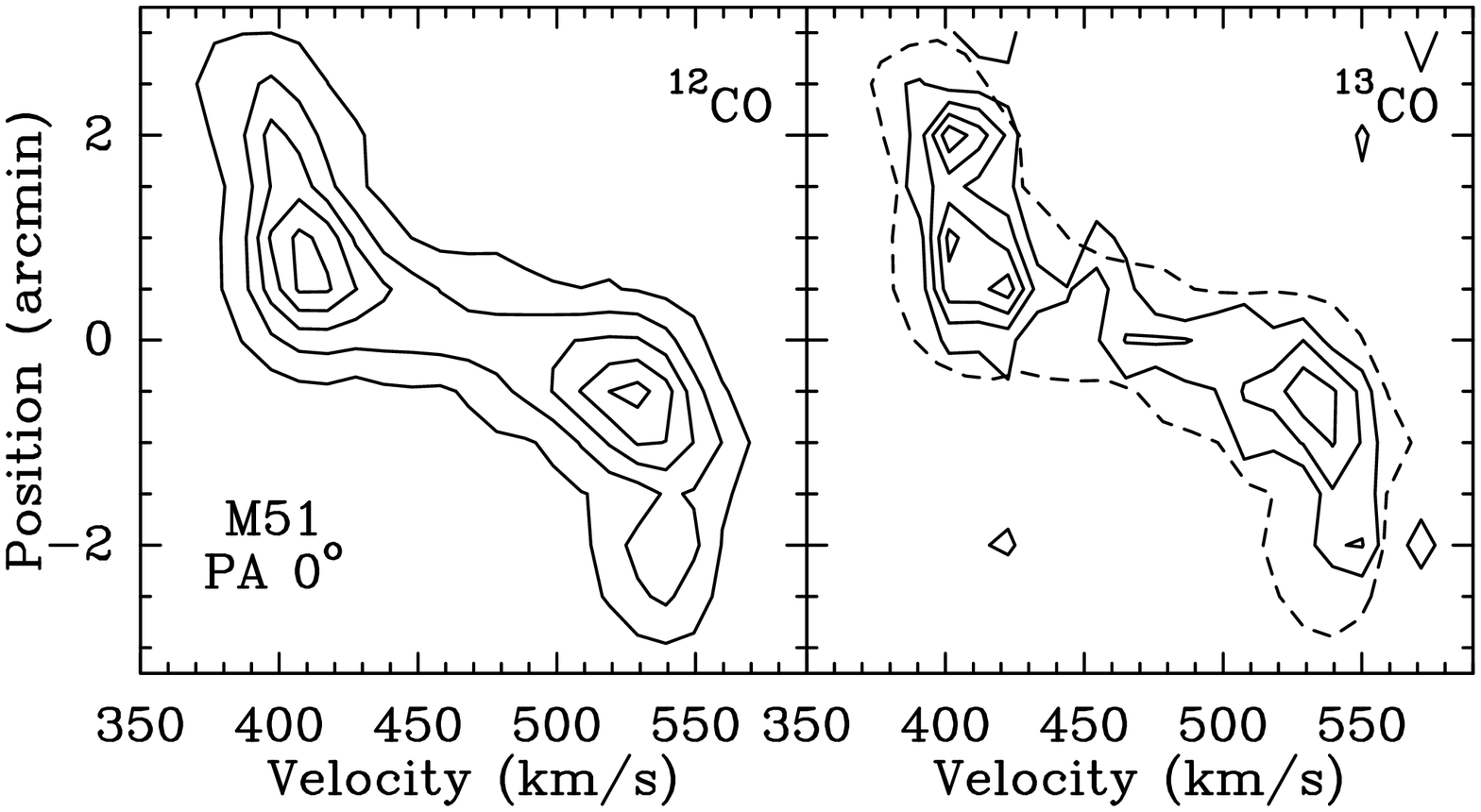}
\end{minipage}%
\begin{minipage}[t]{0.33\textwidth}
\includegraphics[angle=-90,width=\textwidth]{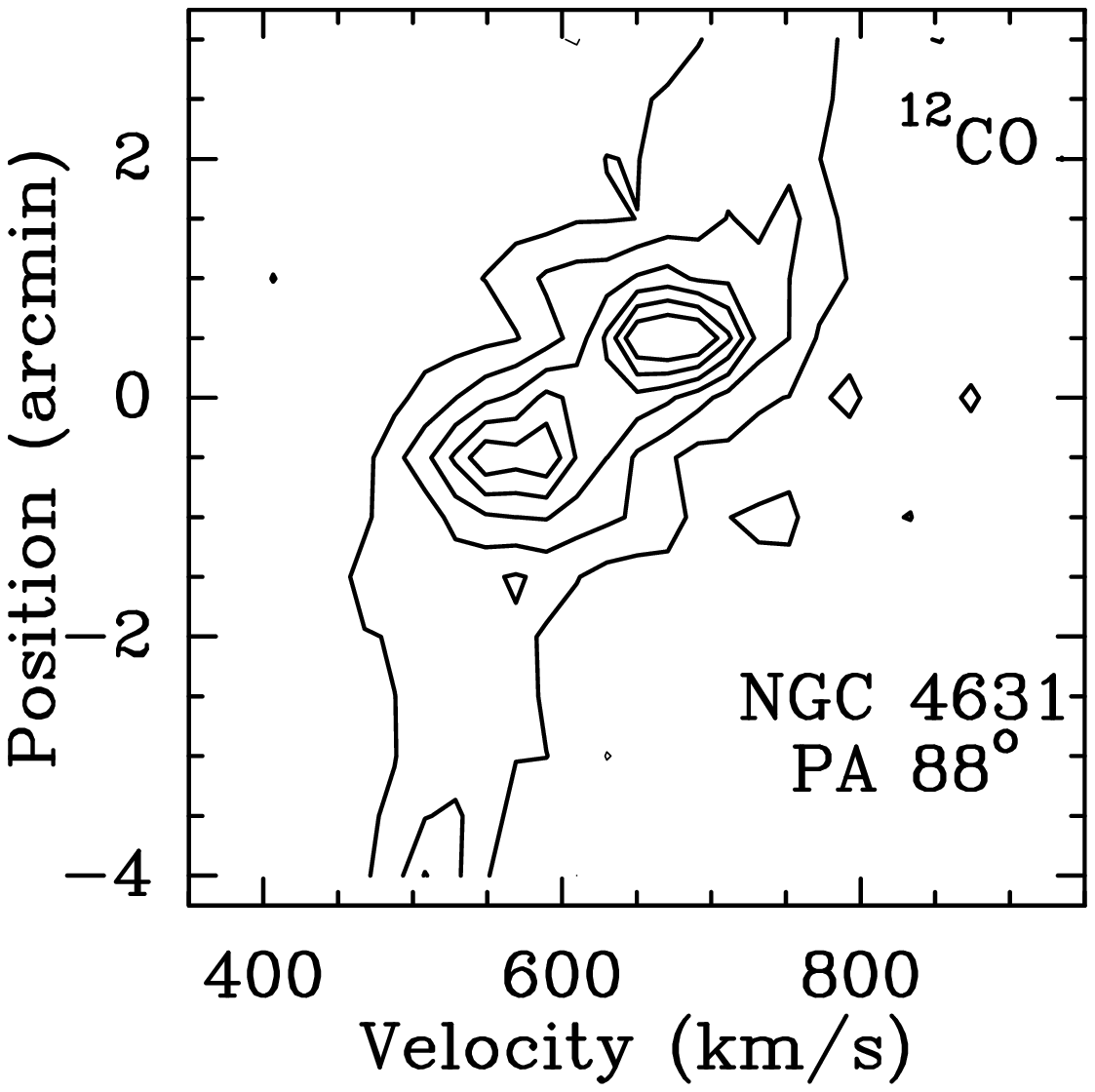}
\end{minipage}

\caption{Position-velocity diagram of $^{12}$CO and $^{13}$CO (only
for NGC~3627 and M51) emission along the major axes of galaxies. For
M82, the $P-V$ diagrams with position angle of 0$^{\circ}$ and
90$^{\circ}$ are shown. All the spectra were smoothed to have a
velocity resolution of $\sim$20\,km s$^{-1}$. The dashed contour on
the $^{13}$CO panel represents the lowest $^{12}$CO contour for
comparison.}\label{Fig4}
\end{figure}


\begin{figure}[htbp]
\centering
\begin{minipage}[t]{0.25\textwidth}
\centering
\includegraphics[width=\textwidth]{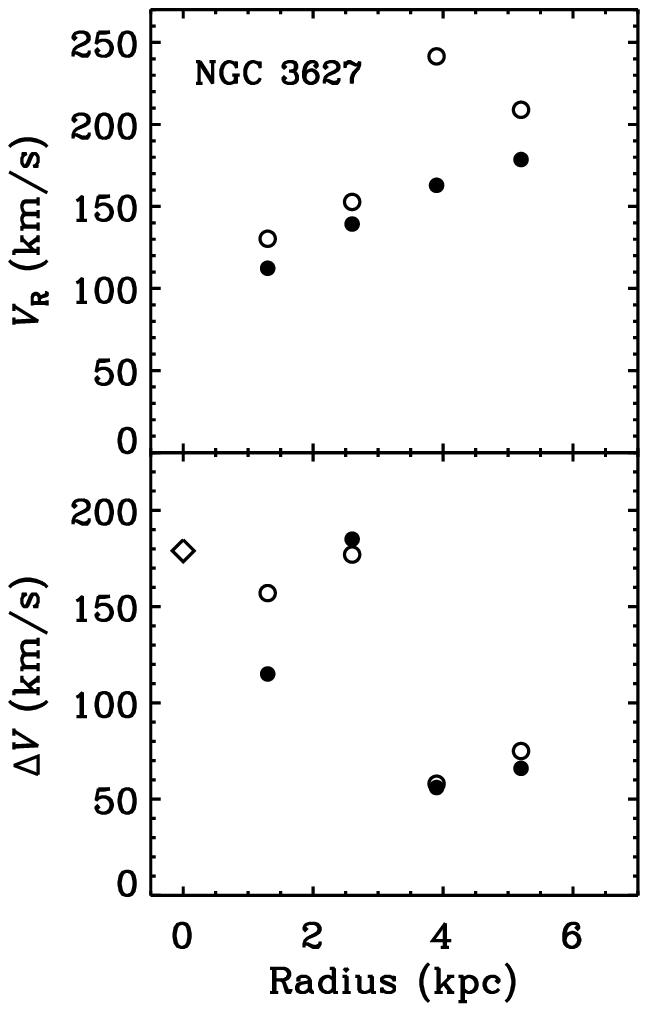}
\end{minipage}%
\begin{minipage}[t]{0.25\textwidth}
\centering
\includegraphics[width=\textwidth]{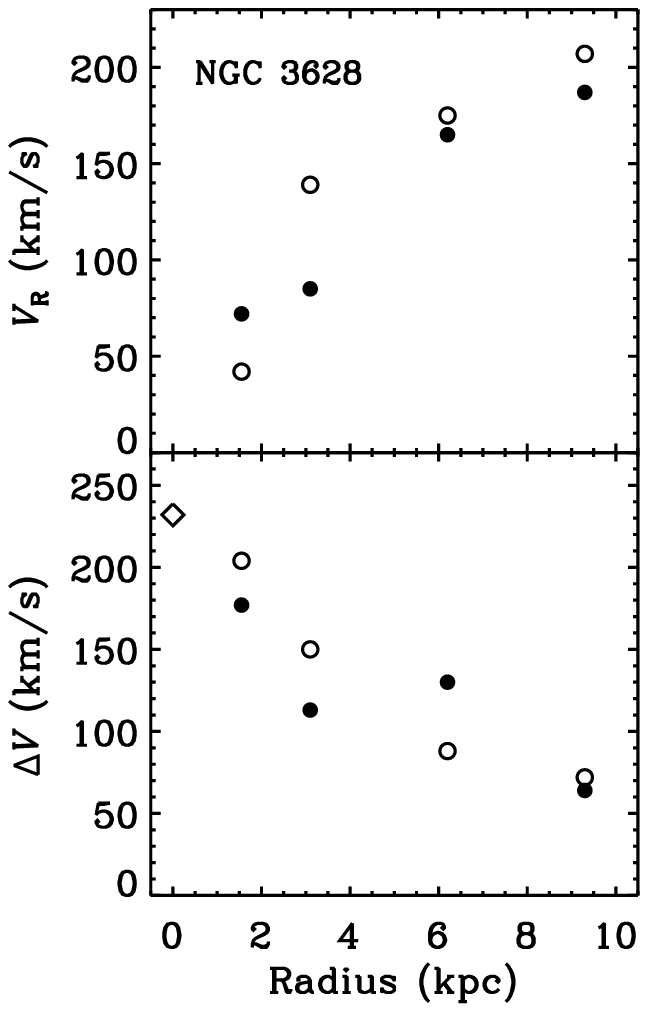}
\end{minipage}%
\begin{minipage}[t]{0.25\textwidth}
\centering
\includegraphics[width=\textwidth]{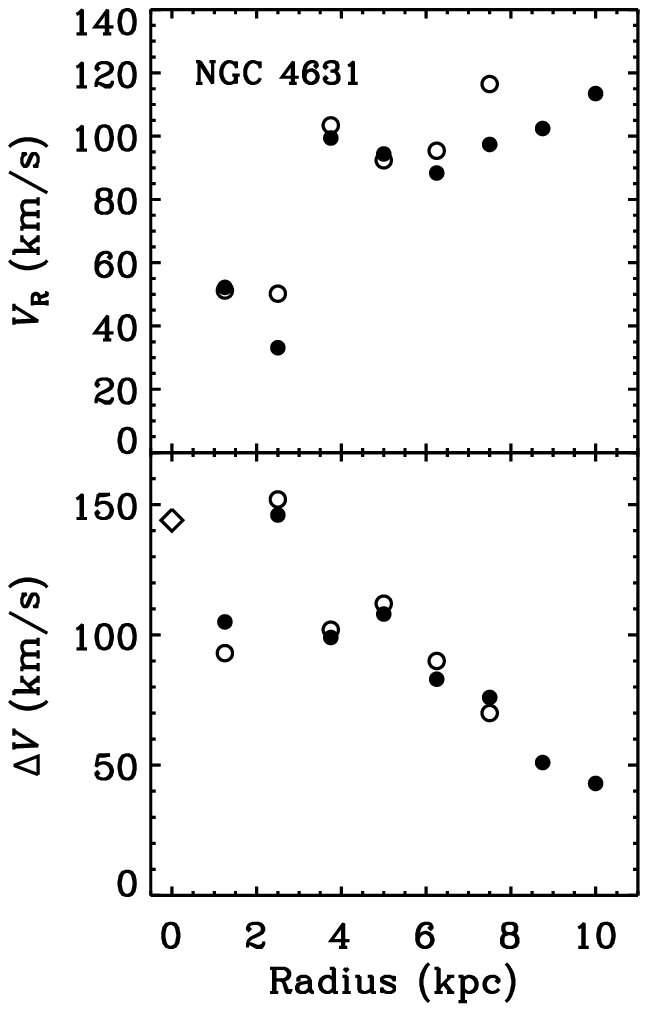}
\end{minipage}%
\begin{minipage}[t]{0.25\textwidth}
\centering
\includegraphics[width=\textwidth]{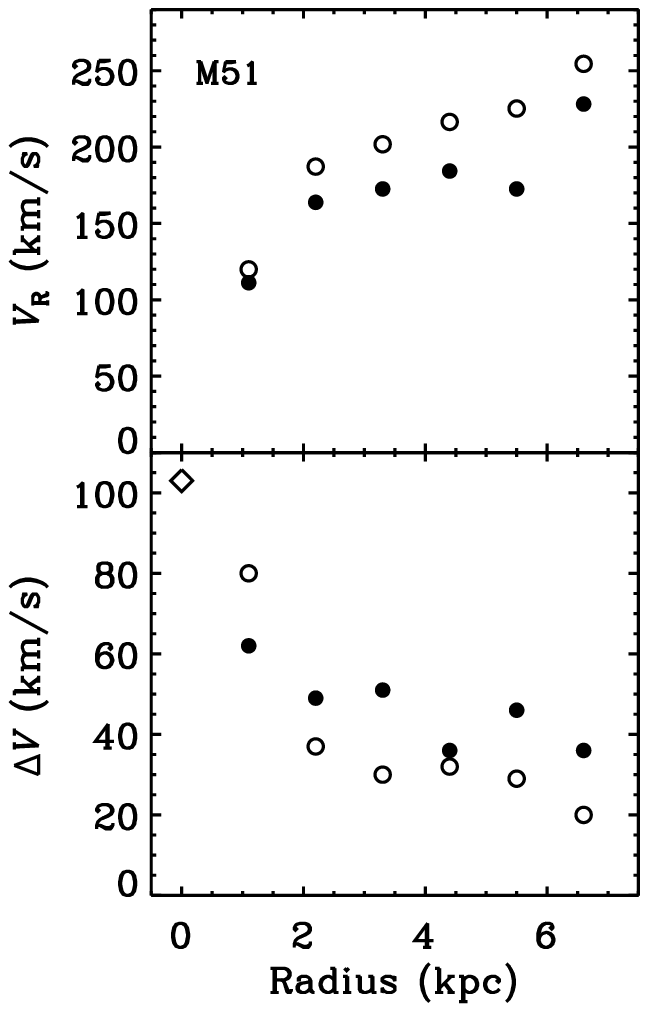}
\end{minipage}

\caption{Rotation velocity derived from \CO\ emission line using
eq.(8) and line width measured at each position along the major
axis. The different symbols represent the same as in
Fig.3}\label{Fig5}
\end{figure}

\begin{figure}[htbp]
\centering
\begin{minipage}[t]{\textwidth}
\centering
\includegraphics[width=\textwidth]{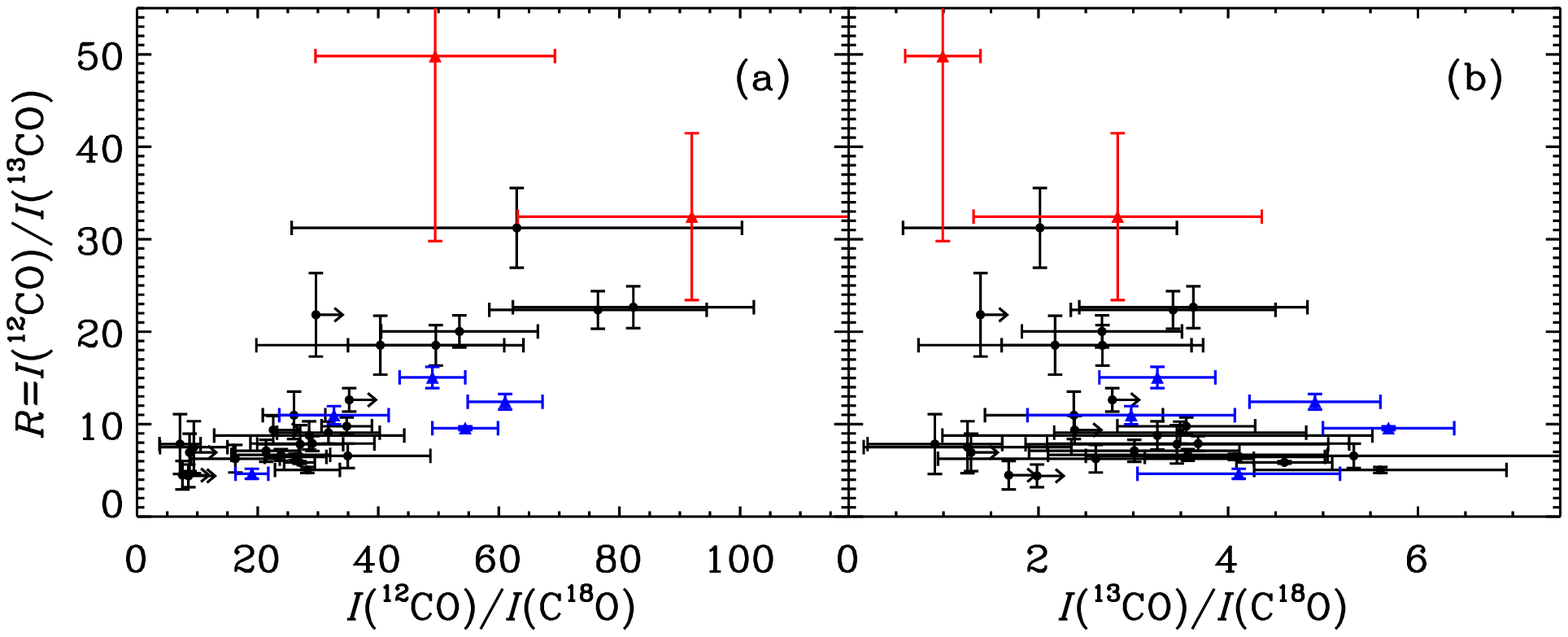}
\end{minipage}\\[10pt]
\begin{minipage}[t]{\textwidth}
\centering
\includegraphics[width=\textwidth]{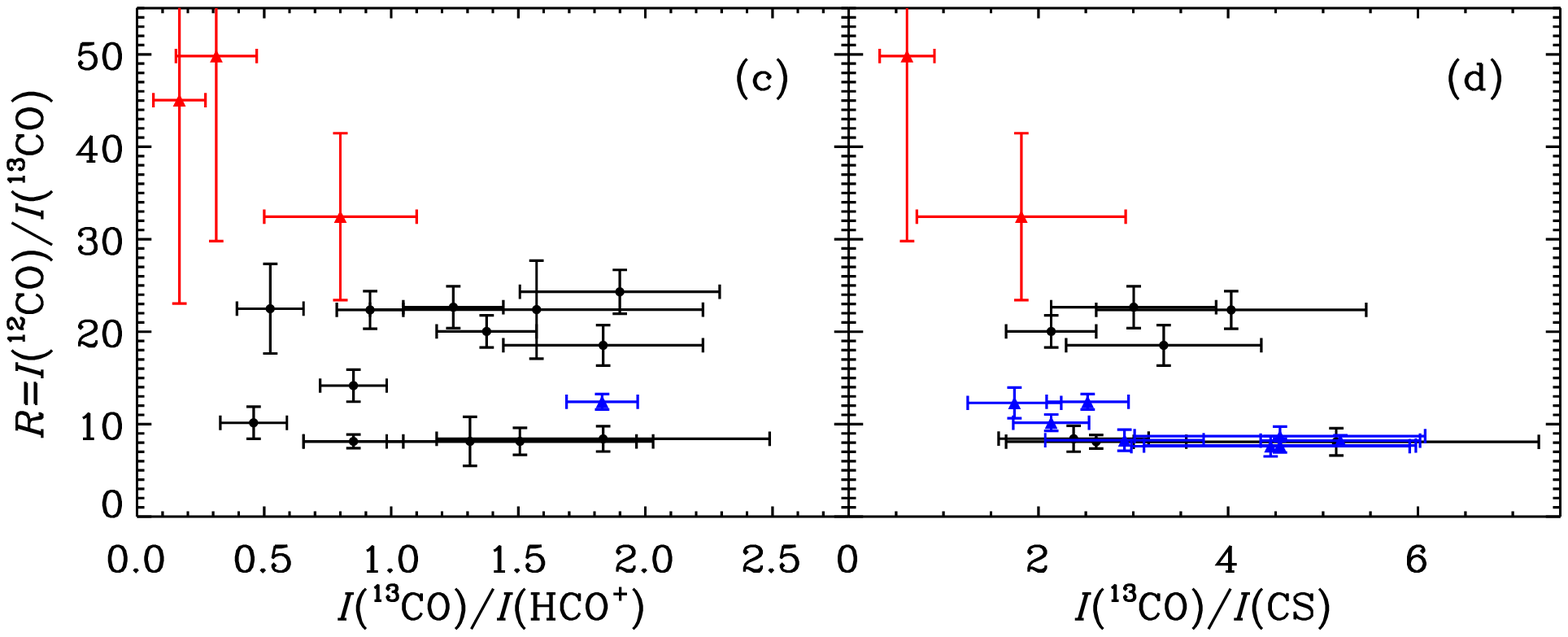}
\end{minipage}
\caption{The relationship for the normal spiral and starburst galaxies between
the integrated intensity ratio of $^{12}$CO/$^{13}$CO and a) $^{12}$CO/C$^{18}$O,
b) $^{13}$CO/C$^{18}$O, c) $^{13}$CO/HCO$^+$, and d) $^{13}$CO/CS. The black
symbol represents the emission of $^{12}$CO, $^{13}$CO, and
C$^{18}$O in M82 and M51 (add some new detections in IC~342 and
NGC~6949); HCO$^+$ and CS in M82, NGC~3628 and M51. Lower limits~(2$\sigma$)
of ratio are denoted with right pointing arrow for some non-detection
C$^{18}$O emission. The triangle symbol represents the data taken from
literatures (see Table~\ref{Table4}), the red triangles: NGC~3256, NGC~6240, Arp~220; the blue
triangles: NGC~253, NGC~1808, NGC~2146, NGC~4826 and Circinus. Note that
all the ratios have been corrected for the different beamsizes.}\label{Fig6}
\end{figure}

\clearpage


\begin{table}
 \bc
\begin{minipage}[c]{65mm}
\caption[c]{Source List and Galaxy Properties}\label{Table1}
\end{minipage}
\scriptsize
\tabcolsep 1mm
\begin{tabular}{lcccclllrrc}
\hline\noalign{\smallskip}
Source  & Alias & R.A. & Decl. & $V$ & $i$ & P.A. & Type & $D_{25}$ & $D$ & $d$ \\
 &  & (J2000.0) & (J2000.0) & (km s$^{-1}$)
& (deg) & (deg) &  & (arcmin) & (Mpc) & (kpc) \\
(1) & (2) & (3) & (4) & (5) & (6) & (7) & (8) & (9) & (10) & (11) \\

\hline\noalign{\smallskip}
NGC 2903 & UGC 5079 & 09 32 10.5 & +21 30 05.0 & 556 & 61 & 17 & SAB(rs)bc, H$\amalg$ & 12.6$\times$6.0 & 7.96 & 1.9  \\
NGC 3031 & M 81 & 09 55 33.6 & +69 03 56.0 & $-$34 & 58 & 157 & SA(s)ab, LINER/Sy1.8 & 26.9$\times$14.1 & 1.00 & 0.3 \\
NGC 3034 & M 82 & 09 55 49.6 & +69 40 41.0 & 203 & 80$^a$ & 65$^a$ & I0, Sbrst/H$\amalg$ & 11.2$\times$4.3 & 2.90 & 1.3  \\
NGC 3521 & UGC 6150 & 11 05 49.2 & $-$00 02 15.0 & 805 & 58 & 164 & SAB(rs)bc, H$\amalg$/LINER & 11.0$\times$5.1 & 11.47 & 2.6  \\
NGC 3627 & M 66 & 11 20 15.3 & +12 59 32.0 & 727 & 63 & 176 & SAB(s)b, LINER/Sy2 & 9.1$\times$4.2 & 10.41  & 2.6  \\
NGC 3628 & UGC 6350 & 11 20 16.2 & +13 35 22.0 & 843 & 89$^a$ & 103$^a$ & SAb pec sp, H$\amalg$/LINER & 14.8$\times$3.0 & 12.07  & 3.1  \\
NGC 4631 & UGC 7865 & 12 42 07.1 & +32 32 33.0 & 606 & 85$^a$ & 88$^a$ & SB(s)d & 15.5$\times$2.7 & 8.67 & 2.5  \\
NGC 4736 & M 94 & 12 50 52.9 & +41 07 15.0 & 308 & 35 & 100 & (R)SA(r)ab, Sy2/LINER & 11.2$\times$9.1 & 4.40 & 1.4  \\
NGC 5055 & M 63 & 13 15 49.5 & +42 01 39.0 & 504 & 55 & 105 & SA(rs)bc, H$\amalg$/LINER & 12.6$\times$7.2 & 7.21 & 2.3  \\
NGC 5194 & M 51a & 13 29 53.5 & +47 11 42.0 & 463 & 20$^a$ & 0 & SA(s)bc pec, H$\amalg$/Sy2.5 & 11.2$\times$6.9 & 6.62 & 2.2  \\
NGC 5457 & M 101 & 14 03 09.0 & +54 21 24.0 & 241 & 27 & 40 & SAB(rs)cd & 28.8$\times$26.9 & 3.45 & 1.4 \\

\noalign{\smallskip}\hline
\end{tabular}
\ec
\tablenotes{a}{\textwidth}{From \cite*{Young95}.}
\tablecomments{\textwidth}{Cols.(1) and (2): Galaxy name.
Col.(3) and (4): Adopted tracking center. Units of right ascension
are hours,minutes,and seconds,and units of declination are
degrees,arcminutes,and arcseconds. Col.(5): Heliocentric velocity
drawn from the literature and NED. Cols.(6) and (7): Inclination
($i$) and position angle (P.A.) from Helfer et al.2003, except where
noted. Col.(8): Morphological type and nuclear classification from
the NED database. Col.(9): Major- and minor- axis diameters from the
NED database. Col.(10): Luminosity distance, calculated assuming
$H_0$ = 70 km s$^{-1}$MPc$^{-1}$. Col.(11): Linear scale of
60${''}$ at distance $D$, taken from the NED database.}

\end{table}

\begin{table}
\bc
\begin{minipage}[c]{85mm}
\caption{Observed and Derived Properties of $^{12}$CO and
$^{13}$CO}\label{Table2}
\end{minipage}

\tabcolsep 0.8mm
\tiny
\begin{tabular}{lrrrcccrccrrrr}
\hline\noalign{\smallskip}
\multicolumn{1}{c}{} &
\multicolumn{1}{c}{} &
\multicolumn{1}{c}{} &
\multicolumn{3}{c}{$^{12}$CO} &
\multicolumn{1}{c}{} &
\multicolumn{3}{c}{$^{13}$CO} &
\multicolumn{1}{c}{} &
\multicolumn{1}{c}{} &
\multicolumn{1}{c}{} &
\multicolumn{1}{c}{} \\
\cline{4-6}
\cline{8-10}
\multicolumn{1}{c}{Source} &
\multicolumn{1}{c}{$\Delta \alpha^a$} &
\multicolumn{1}{c}{$\Delta \delta^a$} &
\multicolumn{1}{c}{$I_{\rm ^{12}CO}\pm \sigma_{I}b$} &
\multicolumn{1}{c}{$V_{\rm mean}^c$} &
\multicolumn{1}{c}{$\Delta V^c$} &
\multicolumn{1}{c}{} &
\multicolumn{1}{c}{$I_{\rm ^{12}CO}\pm \sigma_{I}^b$} &
\multicolumn{1}{c}{$V_{\rm mean}^c$} &
\multicolumn{1}{c}{$\Delta V^c$} &
\multicolumn{1}{c}{${\cal R}^d$} &
\multicolumn{1}{c}{$\tau$($^{13}{\rm CO})^e$} &
\multicolumn{1}{c}{$N({\rm H}_2)^f$}  &
\multicolumn{1}{c}{$T_k^g$} \\
\multicolumn{1}{c}{} &
\multicolumn{1}{c}{(arcmin)} &
\multicolumn{1}{c}{(arcmin)} &
\multicolumn{1}{c}{(K km s$^{-1}$)} &
\multicolumn{1}{c}{(km s$^{-1}$)} &
\multicolumn{1}{c}{(km s$^{-1}$)} &
\multicolumn{1}{c}{} &
\multicolumn{1}{c}{(K km s$^{-1}$)} &
\multicolumn{1}{c}{(km s$^{-1}$)} &
\multicolumn{1}{c}{(km s$^{-1}$)} &
\multicolumn{1}{c}{} &
\multicolumn{1}{c}{} &
\multicolumn{1}{c}{$(10^{20}\ {\rm cm}^{-2})$} &
\multicolumn{1}{c}{(K)} \\
 \hline\noalign{\smallskip}
NGC 2903 & 0.00 & 0.00 & 24.89$\pm$0.92 & 548 & 155 & & 1.95$\pm$0.58 & 591 & 151 & 12.8$\pm$4.3 & 0.08  & 49.8 & 55 \\
NGC 3031 & 0.00 & 0.00 & 0.66$\pm$0.28 & -38 & 127 & & $<0.68$ & ... & ... & $>$1.0 & ...  & 1.3 & ... \\
M82 & 0.00 & 0.00 & 224.61$\pm$1.51 & 159 & 138 &  & 10.25$\pm$0.82 & 126 & 123 & 21.9$\pm$1.9 & 0.05 & 449.2 & 98 \\
         & 0.00 & 0.50 & 163.0$\pm$1.62 & 180 & 162 &  & 11.80$\pm$1.06 & 153 & 167 & 13.8$\pm$1.4 & 0.08 & 328.8 & 66 \\
         & 0.00 & -0.50 & 113.4$\pm$1.12 & 150 & 115 &  & 5.59$\pm$0.90 & 137 & 150 & 20.3$\pm$3.5 & 0.05 & 230.0 & 110 \\
         & 0.50 & 0.00 & 344.77$\pm$2.04 & 214 & 184 &  & 13.91$\pm$1.31 & 230 & 194 & 24.8$\pm$2.5 & 0.04 & 689.5 & 111 \\
         & 0.50 & 0.50 & 278.07$\pm$1.53 & 233 & 183 &  & 11.37$\pm$0.97 & 238 & 180 & 24.5$\pm$2.2 & 0.04 & 556.1 & 110 \\
         & 0.50 & -0.50 & 134.00$\pm$1.50 & 169 & 125& & 5.61$\pm$1.10 & 162 & 87 & 23.9$\pm$4.9 & 0.04 & 269.8 & 110 \\
         & 1.00 & 0.00 & 228.83$\pm$1.66 & 242 & 180 &  & 11.29$\pm$1.25 & 265 & 142 & 20.3$\pm$2.4 & 0.05 & 457.7 & 90 \\
         & 1.00 & 0.50 & 218.00$\pm$1.69 & 268 & 162 &  & 6.38$\pm$0.83 & 313 & 110 & 34.2$\pm$4.7 & 0.03 & 435.8 & 120 \\
     & 1.50 & 0.50 & 92.34$\pm$2.04  & 255 & 164 &  & 6.99$\pm$1.01 & 264 & 215 & 13.2$\pm$2.2 & 0.08 & 184.7 & 57 \\
     & 1.50 & -0.50 & 48.95$\pm$1.74 & 173 & 94 &  & 3.64$\pm$1.36 & 158 & 82 & 13.4$\pm$5.5 & 0.08 & 97.9 & 58 \\
     & 2.50 & 0.00 & 48.64$\pm$2.44  & 219 & 147 &  & 1.94$\pm$0.96 & 158 & 46 & 25.1$\pm$13.7 & 0.04 & 97.3 & 113 \\
     & -0.50 & 0.00 & 75.91$\pm$1.88 & 144 & 112 &  & 6.78$\pm$2.83 & 114 & 100 & 11.2$\pm$5.0 & 0.09 & 151.8 & 48 \\
     & -0.50 & -0.50 & 45.75$\pm$1.66 & 140 & 110 & & 5.75$\pm$1.33 & 226 & 250 & 7.9$\pm$2.1 & 0.13 & 91.5 & 32 \\
     & -1.00 & 0.00 & 47.85$\pm$1.96 & 135 & 104 & & 2.52$\pm$1.49 & 141 & 44 & 18.9$\pm$12.0 & 0.05 & 95.7 & 84 \\
     & -1.50 & 0.00 & 47.40$\pm$3.86 & 162 & 147 & & 5.81$\pm$2.16 & 183 & 111 & 8.2$\pm$3.7 & 0.13 & 94.8 & 33 \\
NGC 3521 & 0.00 & 0.00 & 22.00$\pm$0.37 & 770 & 231 &  & 2.05$\pm$0.66 & 770 & 210 & 10.7$\pm$3.7 & 0.10 & 44.0 & 45\\
NGC 3627 & 0.00 & 0.00 & 17.24$\pm$0.87 & 769 & 179 &  & 1.66$\pm$0.49 & 866 & 200 & 10.4$\pm$3.6 & 0.10 & 34.5 & 44 \\
         & 0.00 & 0.50 & 25.26$\pm$0.83 & 653 & 157 &  & 4.06$\pm$0.98 & 625 & 163 & 6.2$\pm$1.7 & 0.18 & 50.5 & 24 \\
         & 0.00 & 1.00 & 13.88$\pm$0.38 & 633 & 177 &  & 2.28$\pm$0.51 & 597 & 138 & 6.1$\pm$1.5 & 0.18 & 27.8 & 23 \\
         & 0.00 & 1.50 & 8.55$\pm$0.52 & 554 & 58 &  & 1.46$\pm$0.48 & 580 & 75 & 5.9$\pm$2.3 & 0.19 & 17.1 & 22 \\
         & 0.00 & 2.00 & 6.90$\pm$0.50 & 583 & 75 &  & 1.79$\pm$0.35 & 527 & 74 & 3.9$\pm$1.0 & 0.30 & 13.8 & 13 \\
         & 0.00 & -0.50 & 22.25$\pm$1.96 & 827 & 115&  & 2.87$\pm$0.69 & 829 & 145 & 7.7$\pm$2.5 & 0.14 & 44.5 & 31 \\
         & 0.00 & -1.00 & 8.96$\pm$0.53 & 851 & 185 &  & 1.78$\pm$0.35 & 885 & 91 & 5.0$\pm$1.3 & 0.22 & 17.9 & 18 \\
         & 0.00 & -1.50 & 6.82$\pm$0.31 & 872 & 56 &  & 1.96$\pm$0.41 & 877 & 90  & 3.5$\pm$0.9 & 0.34 & 13.6 & 11 \\
         & 0.00 & -2.00 & 5.90$\pm$0.42 & 886 & 66 &  & 1.02$\pm$0.40 & 902 & 43 & 5.1$\pm$2.5 & 0.19 & 11.8 & 22 \\
NGC 3628 & 0.00 & 0.00 & 35.80$\pm$0.67 & 823 & 232 &  & 3.22$\pm$0.50 & 839 & 211 & 11.1$\pm$1.9 & 0.09 & 71.6 & 47 \\
         & 0.49 & -0.11 & 28.95$\pm$0.52 & 865 & 204 &  & $<$2.52 & ... & ... & $>$11.5 & ... & 57.9 & ... \\
         & 0.98 & -0.22 & 9.95$\pm$0.77 & 962 & 150 &  & 2.52$\pm$0.60 & 986 & 110 & 3.9$\pm$1.2 & 0.29 & 19.9 & 13 \\
         & 1.96 & -0.44 & 3.77$\pm$0.92 & 998 & 88 &  & $<$1.76 & ... & ... & $>$2.1 & ... & 7.5 & ... \\
         & 2.94 & -0.66 & 2.13$\pm$0.64 & 1030 & 72 &  & $<$1.18 & ... & ... & $>$1.8 & ... & 4.3 & ...  \\
         & -0.49 & 0.11 & 14.06$\pm$0.46 & 751 & 177 &  & 3.07$\pm$0.49 & 745 & 206 & 4.6$\pm$0.9 & 0.25 & 28.1 & 16 \\
         & -0.98 & 0.22 & 16.16$\pm$1.47 & 738 & 113 &  & $<$2.04 & ... & ... & $>$7.9 & ... & 32.3 & ... \\
         & -1.96 & 0.44 & 6.45$\pm$0.81 & 658 & 130 &  & 1.35$\pm$0.48 & 675 & 83 & 4.8$\pm$2.3 & 0.23 & 12.9 & 17 \\
         & -2.94 & 0.66 & 2.95$\pm$1.11 & 636 & 64 &  & $<$2.22 & ... & ... & $>$1.3 & ... & 5.9 & ... \\
NGC 4631 & 0.00 & 0.00 & 17.36$\pm$0.62 & 623 & 144 &  & 1.65$\pm$0.46 & 606 & 146 & 10.5$\pm$3.3 & 0.10 & 34.7 & 44 \\
         & 0.51 & 0.02 & 23.77$\pm$0.24 & 674 & 93 &  & 2.53$\pm$0.46 & 677 & 80 & 9.4$\pm$1.8 & 0.11 & 47.5 & 39 \\
         & 1.00 & 0.03 & 13.82$\pm$0.67 & 673 & 152 &  & 1.48$\pm$0.42 & 697 & 198 & 9.3$\pm$3.0 & 0.11 & 27.6 & 39 \\
         & 1.50 & 0.06 & 4.98$\pm$0.26 & 726 & 102 &  & 0.71$\pm$0.27 & 704 & 108 & 7.0$\pm$3.0 & 0.15 & 9.9 & 28 \\
         & 2.00 & 0.06 & 4.07$\pm$0.25 & 715 & 112 &  & 1.25$\pm$0.21 & 731 & 148 & 3.3$\pm$0.7 & 0.37 & 8.1 & 10 \\
         & 2.50 & 0.11 & 3.29$\pm$0.19 & 718 & 90 &  & $<$0.52 & ... & ... & $>$6.3 & ... & 6.6 & ... \\
         & 2.99 & 0.22 & 2.48$\pm$0.33 & 739 & 70 &  & $<$0.80 & ... & ... & $>$3.1 & ... & 5.0 & ... \\
         & -0.51 & -0.02 & 19.50$\pm$0.27 & 571 & 105 &  & 1.76$\pm$0.31 & 528 & 141 & 11.1$\pm$2.1 & 0.09 & 39.0 & 47 \\
         & -1.00 & -0.03 & 11.40$\pm$0.44 & 590 & 146 &  & 1.77$\pm$0.47 & 600 & 140 & 6.4$\pm$1.9 & 0.17 & 22.8 & 25 \\
         & -1.50 & -0.06 & 3.41$\pm$0.23 & 524 & 99 &  & $<$0.83 & ... & ... & $>$4.0 & ... & 6.8 & ... \\
         & -2.00 & -0.07 & 1.89$\pm$0.16 & 529 & 108 &  & $<$0.52 & ... & ... & $>$3.6 & ... & 3.8 & ... \\
         & -2.50 & -0.11 & 1.96$\pm$0.14 & 535 & 83 &  & $<$0.40 & ... & ... & $>$4.9 & ... & 3.9 & ... \\
         & -2.99 & -0.22 & 2.20$\pm$0.31 & 526 & 76 &  & $<$0.90 & ... & ... & $>$2.4 & ... & 4.4 & ... \\
         & -3.50 & -0.38 & 2.86$\pm$0.19 & 521 & 51 &  & $<$1.10 & ... & ... & $>$2.6 & ... & 5.7 & ... \\
         & -3.68 & -0.30 & 3.68$\pm$0.30 & 510 & 43 &  & 0.71$\pm$0.25 & 486 & 34 & 5.2$\pm$2.2 & 0.21 & 7.4 & 19 \\
NGC 4736 & 0.00 & 0.00 & 8.01$\pm$0.70 & 340 & 153 &  & 1.12$\pm$0.29 & 391 & 114 & 7.2$\pm$2.5 & 0.15 & 16.0 & 28 \\
NGC 5055 & 0.00 & 0.00 & 16.15$\pm$1.05 & 527 & 214 &  & 2.09$\pm$0.86 & 527 & 275 & 7.7$\pm$3.7 & 0.14 & 32.3 & 31 \\
M51 & 0.00 & 0.00 & 33.35$\pm$0.39 & 475 & 103 &  & 3.86$\pm$0.35 & 472 & 87 & 8.6$\pm$0.9 & 0.12 & 66.7 & 35 \\
         & 0.00 & 0.50 & 28.90$\pm$0.42 & 434 & 80 &  & 3.01$\pm$0.48 & 425 & 37 & 9.6$\pm$1.7 & 0.11 & 57.8 & 40 \\
         & 0.00 & 1.00 & 17.45$\pm$0.56 & 410 & 30 &  & 2.43$\pm$0.41 & 408 & 23 & 7.2$\pm$1.4 & 0.15 & 34.9 & 29 \\
         & 0.00 & 1.50 & 10.35$\pm$0.50 & 406 & 30 &  & 1.26$\pm$0.41 & 409 & 41 & 8.2$\pm$3.1 & 0.13 & 20.7 & 34 \\
         & 0.00 & 2.00 & 9.70$\pm$0.45 & 401 & 32 &  & 1.98$\pm$0.59 & 406 & 26 & 4.9$\pm$1.7 & 0.23 & 19.4 & 18 \\
         & 0.00 & 2.50 & 6.90$\pm$0.37 & 398 & 29 &  & 0.80$\pm$0.29 & 396 & 37 & 8.6$\pm$3.5 & 0.12 & 13.8 & 36 \\
         & 0.00 & 3.00 & 2.62$\pm$0.56 & 388 & 20 &  & $<$1.00 & ... & ... & $>$2.6 & ... & 5.2 & ... \\
         & 0.00 & -0.50 & 27.35$\pm$0.53 & 513 & 62 &  & 2.28$\pm$0.49 & 520 & 58 & 12.0$\pm$2.8 & 0.09 & 54.7 & 51 \\
         & 0.00 & -1.00 & 16.25$\pm$0.55 & 534 & 49 &  & 1.90$\pm$0.44 & 536 & 42 & 8.5$\pm$2.3 & 0.12 & 32.5 & 35 \\
         & 0.00 & -1.50 & 6.55$\pm$0.23 & 534 & 51 &  & 0.86$\pm$0.22 & 563 & 53 & 7.6$\pm$2.2 & 0.14 & 13.1 & 31 \\
         & 0.00 & -2.00 & 7.90$\pm$0.37 & 538 & 36 &  & 1.64$\pm$0.38 & 539 & 43 & 4.8$\pm$1.3 & 0.23 & 15.8 & 17 \\
         & 0.00 & -2.50 & 5.77$\pm$0.33 & 534 & 46 &  & $<$0.68 & ... & ... & $>$8.5 & ... & 11.5 & ... \\
         & 0.00 & -3.00 & 2.28$\pm$0.35 & 553 & 36 &  & $<$0.62 & ... & ... & $>$3.7 & ... & 4.6 & ... \\
M51 disk$^h$ &\ldots & \ldots & 11.85$\pm$0.30 & 462 & 151 & & 1.73$\pm$0.37 & 411 & 20 & 6.8$\pm$1.6 & 0.16 & 23.7 & 27 \\
NGC 5457 & 0.00 & 0.00 & 9.82$\pm$0.51 & 255 & 72 &  & 1.97$\pm$0.49 & 277 & 128 & 5.0$\pm$1.5 & 0.22 & 19.6 & 18 \\

\noalign{\smallskip}\hline
\end{tabular}
\ec
\tablenotes{a}{\textwidth}{Offset from the nucleus position listed in Table~\ref{Table1}, in units of arcminutes.}
\tablenotes{b}{\textwidth}{The measured integrated intensities and associated uncertainties, calculated using the prescription explained in the text.
For non-detections, a 2 $\sigma_{I}$ upper limit was given.}\\[14pt]
\tablenotes{c}{\textwidth}{Velocity and line widths are Gaussian fit values, or else are calculated from the moment for non-Gaussian lines.}
\tablenotes{d}{\textwidth}{The ratio of $^{12}$CO to $^{13}$CO integrated intensities. The errors are based on the statistical
uncertainties of integrated intensities, which can be derived from the error transfer formula
$\sigma (\rat_{12/13})=([\frac{\sigma(I_{12})}{I_{13}}]^2+[\frac{\sigma(I_{13})\times\rat_{12/13}}{I_{13}}]^2)^{1/2}$.}\\[14pt]
\tablenotes{e}{\textwidth}{The average optical depth in $^{13}$CO emission line, calculated from eq.(4).}
\tablenotes{f}{\textwidth}{H$_2$ column density calculated from $^{12}$CO data by adopting Galactic standard
$X$-factor, $2.0\times10^{20}$ cm$^{-2}{\rm K}^{-1}{\rm km}^{-1}$s.}\\[14pt]
\tablenotes{g}{\textwidth}{The gas kinetic temperature calculated from the assumption of LTE conditions by equating the column
densities derived from both $^{12}$CO and $^{13}$CO (see text \S~\ref{ss:parameters}).}
\tablenotes{h}{\textwidth}{M51 disk represents the average intensity over the disk region except the nucleus.}

\end{table}


\begin{table}
\bc
\begin{minipage}[c]{75mm}
\caption[c]{Observed and Derived Properties of C$^{18}$O}\label{Table3}
\end{minipage}
\tabcolsep 0.8mm
\scriptsize
\begin{tabular}{lrrccccr}
\hline\noalign{\smallskip}

\multicolumn{1}{c}{Source} &
\multicolumn{1}{c}{$\Delta \alpha$} &
\multicolumn{1}{c}{$\Delta \delta$} &
\multicolumn{1}{c}{$I_{\rm C^{18}O}\pm \sigma_{I}$} &
\multicolumn{1}{c}{$V_{\rm mean}$} &
\multicolumn{1}{c}{$\Delta V$} &
\multicolumn{1}{c}{$\tau(C^{18}O)^a$} &
\multicolumn{1}{c}{$N(\rm H_2)^b$} \\
\multicolumn{1}{c}{Source} &
\multicolumn{1}{c}{arcmin} &
\multicolumn{1}{c}{arcmin} &
\multicolumn{1}{c}{(K km s$^{-1}$)} &
\multicolumn{1}{c}{(km s$^{-1}$)} &
\multicolumn{1}{c}{(km s$^{-1}$)} &
\multicolumn{1}{c}{} &
\multicolumn{1}{c}{$(10^{21}\ {\rm cm}^{-2})$}  \\
\hline\noalign{\smallskip}
M82 & 0.0 & 0.0  & 3.81$\pm$0.90 & 140 & 121 & 0.017 & 18.6  \\
    & 0.0 & 0.5  & $<$4.2 & \ldots & \ldots & \ldots & \ldots \\
    & 0.0 &-0.5  & 2.55$\pm$1.27 & 224 & 176 & 0.022 & 14.0 \\
    & 0.5 & 0.0  & 3.80$\pm$0.90 & 145 & 93 & 0.011 & 25.2  \\
    & 0.5 & 0.5  & 3.30$\pm$0.76 & 289 & 135 & 0.012 & 20.5  \\
    & 0.5 &-0.5  & $<$4.1 & \ldots & \ldots & \ldots & \ldots \\
    & 1.0 & 0.0  & 4.19$\pm$1.20 & 284 & 236 & 0.018 & 20.6 \\
    & 1.0 & 0.5 & 3.14$\pm$1.84 & 256 & 174 & 0.015 & 14.8 \\
M51 & 0.0 & 0.0 & 1.04$\pm$0.35 & 466 & 176 & 0.032 & 7.4  \\
         & 0.0 & 0.5 & 0.92$\pm$0.49 & 413 & 154 & 0.032 & 5.7  \\
         & 0.0 & 1.0 & 0.45$\pm$0.16 & 407 & 23 & 0.026 & 4.6  \\
     & 0.0 & 1.5 & 0.99$\pm$0.54 & 393 & 79 & 0.10 & 2.4 \\
     & 0.0 & 2.0 & $<$1.17 & \ldots & \ldots & \ldots & \ldots \\
     & 0.0 & 2.5 & 0.88$\pm$0.37 & 369 & 58 & 0.14 & 1.6 \\
         & 0.0 &-0.5 & 0.95$\pm$0.17 & 519 & 37 & 0.035 & 4.3 \\
         & 0.0 &-1.0 & 0.55$\pm$0.13 & 544 & 35 & 0.034 & 3.6  \\
     & 0.0 &-1.5 & $<$0.68 & \ldots & \ldots & \ldots & \ldots \\
     & 0.0 &-2.0 & $<$0.84 & \ldots & \ldots & \ldots & \ldots \\
\noalign{\smallskip}\hline
\end{tabular}
\ec
\tablenotes{a}{\textwidth}{The average optical depth in C$^{18}$O emission
line is calculated from the similar equation as eq.(4).}
\tablenotes{b}{\textwidth}{H$_2$ column density derived from C$^{18}$O, calculated from eq.(6).}

\end{table}
\begin{table}
\bc
\begin{minipage}[c]{55mm}
\caption{Line Intensity Ratios}\label{Table4}
\end{minipage}
\tabcolsep 0.8mm
\scriptsize
\begin{tabular}{lrrccccc}
\hline\noalign{\smallskip}
\multicolumn{1}{c}{Source} &
\multicolumn{1}{c}{$\Delta \alpha$} &
\multicolumn{1}{c}{$\Delta \delta$} &
\multicolumn{1}{c}{$I_{\rm ^{12}CO}/I_{\rm C^{18}O}$} &
\multicolumn{1}{c}{$I_{\rm ^{13}CO}/I_{\rm C^{18}O}$} &
\multicolumn{1}{c}{$I_{\rm ^{13}CO}/I_{\rm HCO^+}$} &
\multicolumn{1}{c}{$I_{\rm ^{13}CO}/I_{\rm CS}$} &
\multicolumn{1}{c}{References} \\
\multicolumn{1}{c}{} &
\multicolumn{1}{c}{arcmin} &
\multicolumn{1}{c}{arcmin} &
\multicolumn{1}{c}{} &
\multicolumn{1}{c}{} &
\multicolumn{1}{c}{} &
\multicolumn{1}{c}{} &
\multicolumn{1}{c}{} \\
\hline\noalign{\smallskip}

M82 & 0.0 & 0.0   & 53.5$\pm$13.0 & 2.7$\pm$0.9 & 1.4$\pm$0.2 & 2.1$\pm$0.5 & This work \\
    & 0.0 & 0.5   & $>$35.3 & $>$2.8 & 0.9$\pm$0.1 & ...         & This work \\
    & 0.0 & -0.5  & 40.5$\pm$20.6 & 2.2$\pm$1.4 & 0.5$\pm$0.1 & ...         & This work \\
    & 0.5 & 0.0   & 82.3$\pm$19.5 & 3.7$\pm$1.2 & 1.2$\pm$0.2 & 3.0$\pm$0.9 & This work \\
    & 0.5 & 0.5   & 76.5$\pm$18.0 & 3.4$\pm$1.1 & 0.9$\pm$0.1 & 4.0$\pm$1.4 & This work \\
    & 0.5 & -0.5  & $>$29.8 & $>$1.4 & 1.6$\pm$0.7 & ...         & This work \\
    & 1.0 & 0.0   & 49.5$\pm$14.5 & 2.7$\pm$1.1 & 1.8$\pm$0.4 & 3.3$\pm$1.0 & This work \\
    & 1.0 & 0.5   & 63.2$\pm$37.5 & 2.0$\pm$1.4 & 1.9$\pm$0.4 & ...         & This work \\
NGC 3628 & 0.0 & 0.0 & ...        & ...         & 0.5$\pm$0.1 & ...         & This work \\
NGC 4631 & 0.0 & 0.0 & $>$22.8 & $>$2.4 & \ldots & \ldots & This work \\
M51      & 0.0 & 0.0 & 29.1$\pm$10.3 & 3.7$\pm$1.6 & 0.9$\pm$0.2 & 2.6$\pm$0.9 & This work \\
         & 0.0 & 0.5 & 28.6$\pm$15.8 & 3.3$\pm$2.3 & 1.8$\pm$0.7 & 2.4$\pm$0.8 & This work \\
         & 0.0 & 1.0 & 34.9$\pm$13.7 & 5.3$\pm$2.8 & 1.3$\pm$0.7 & ...         & This work \\
     & 0.0 & 1.5 & 9.4$\pm$5.6 & 1.3$\pm$1.1 & ... & ... & This work \\
     & 0.0 & 2.0 & $>$7.6 & $>$1.7 & ... & ... & This work \\
     & 0.0 & 2.5 & 7.1$\pm$3.4 & 0.9$\pm$0.7 & \ldots & \ldots & This work \\
     & 0.0 &-0.5 & 26.0$\pm$5.2 & 2.4$\pm$0.9 & 1.5$\pm$0.5 & 5.1$\pm$2.1 & This work \\
     & 0.0 &-1.0 & 27.0$\pm$7.1 & 3.5$\pm$1.6 & ... & ...  & This work \\
     & 0.0 &-1.5 & $>$8.7 & $>$1.3 & \ldots & \ldots & This work \\
     & 0.0 &-2.0 & $>$8.6 & $>$2.0 & \ldots & \ldots & This work \\
M51 disk & \ldots & \ldots & 16.4$\pm$7.4 & 2.6$\pm$1.7 & \ldots & \ldots & This work \\
NGC 253  & 0.0 & 0.0 & 61.0$\pm$6.2 & 4.9$\pm$0.7 & 1.8$\pm$0.1 & 2.5$\pm$0.4 & 1,2,3 \\
NGC 891  & 0.0 & 0.0 & ...          & ...         & ...         & 4.5$\pm$1.4 & 3,4 \\
NGC 1068 & 0.0 & 0.0 & ...          & ...         & ...         & 1.7$\pm$0.5 & 3,5 \\
NGC 1808 & 0.0 & 0.0 & 48.9$\pm$5.4 & 3.3$\pm$0.6 & ...         & ...         & 6,7 \\
NGC 2146 & 0.0 & 0.0 & 32.7$\pm$9.1 & 3.0$\pm$1.1 & ...         & ...         & 6,7 \\
NGC 2903 & 0.0 & 0.0 & ...          & ...         & ...         & 4.5$\pm$1.5 & 3,4 \\
NGC 3256 & 0.0 & 0.0 & 91.6$\pm$28.1 & 2.9$\pm$1.5 & 0.8$\pm$0.3 & 1.8$\pm$1.1 & 8  \\
NGC 4736 & 0.0 & 0.0 & ...          & ...         & ...         & 4.5$\pm$1.5 & 3,4 \\
NGC 4826 & 0.0 & 0.0 & 19.0$\pm$2.7 & 4.1$\pm$1.1 & ...         & ...         & 6,7 \\
NGC 5457 & 0.0 & 0.0 & ...          & ...         & ...         & 2.9$\pm$0.8 & 3,4 \\
NGC 6240 & 0.0 & 0.0 & ...          & ...         & 0.2$\pm$0.1 & ...         & 9 \\
Maffei2  & 0.0 & 0.0 & ...          & ...         & ...         & 5.2$\pm$0.8 & 3,4 \\
IC 342   & 0.0 & 0.0 & 35.1$\pm$3.0 & 3.9$\pm$0.3 & ...         & 2.1$\pm$0.4 & 1,3,4 \\
Circinus & 0.0 & 0.0 & 54.4$\pm$5.4 & 5.7$\pm$0.7 & ...         & ...         & 6,7 \\
Arp 220  & 0.0 & 0.0 & 49.4$\pm$19.9 & 1.0$\pm$0.4 & 0.3$\pm$0.1 & 0.6$\pm$0.3 & 9 \\

\noalign{\smallskip}\hline
\end{tabular}
\ec
\tablecomments{\textwidth}{All the ratios of integrated
intensity have been corrected for the different beamsizes.}
\tablerefs{\textwidth}{(1) \cite*{Sage91b}; (2) \cite*{Henkel93}; (3)
\cite*{Sage90}; (4) \cite*{Sage&Isbell91}; (5) \cite*{Young86};
(6) \cite*{Aalto91}; (7) \cite*{Aalto95};
(8) \cite*{Casoli92}; (9) \cite*{Greve09}.}

\end{table}
\clearpage
\appendix

\section{The Stability of DLH Radio Telescope}
\label{stability}
\begin{figure}[htbp]
\centering
\begin{minipage}[t]{0.5\textwidth}
\centering
\includegraphics[width=\textwidth]{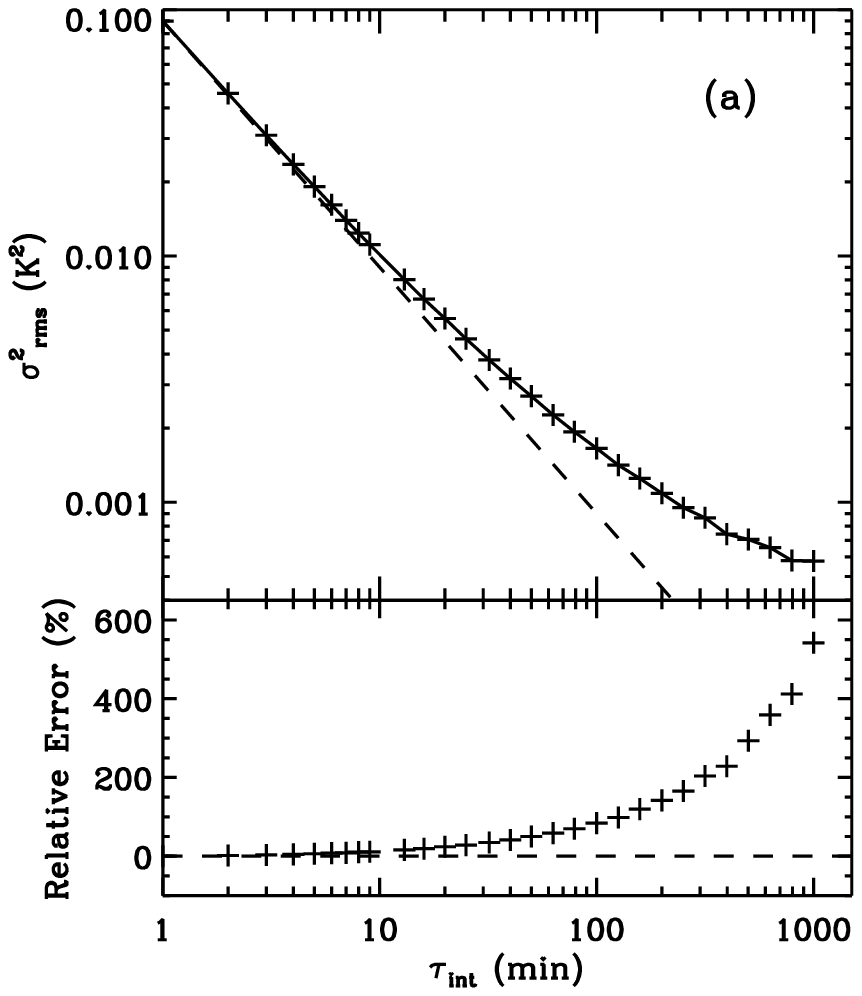}
\end{minipage}%
\begin{minipage}[t]{0.5\textwidth}
\centering
\includegraphics[width=\textwidth]{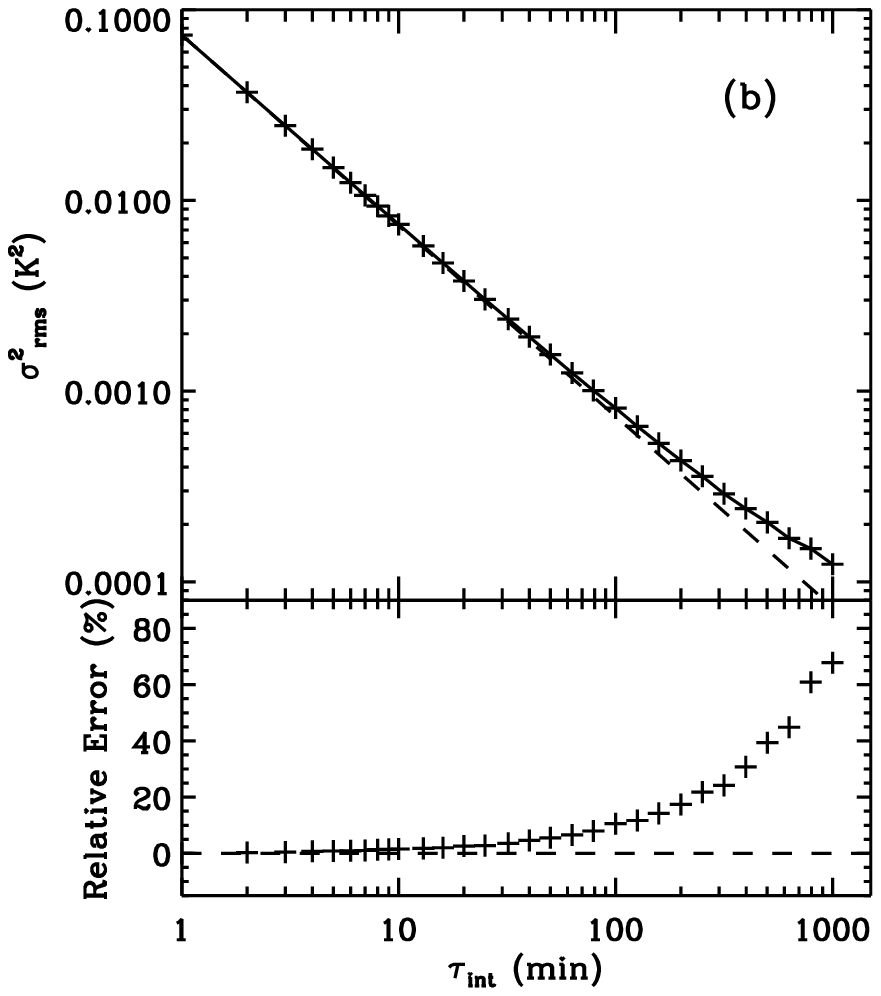}
\end{minipage}

\caption{The Allan Variance and the relative error from radiometer
equation as a function of integration time on source (see
Appendix~\ref{stability}). On the upper panel of each figure, the
dashed line represents what is expected from the radiometer equation
with a slope of -1. On the lower panel, each point represent the relative
error between the value of measured and expected from radiometer equation.
(a)the data taken from the observing semester of 2008 to 2009. (b)the
data taken from the observing semester of 2009 to 2010. }\label{Fig7}
\end{figure}

Figure~\ref{Fig7} is an Allan Variance Plot, which
is often the ultimate way to measure stability but requires an
enormous amount of observation time. There are three main
contributions to be aware of in the Allan plot, including the white
noise, the 1/$f$-noise and low frequency drift noise. The upper
panel of Fig.~\ref{Fig7} shows that how the squared RMS noise of \CO\
spectra with velocity resolution of $\sim$0.16\,km s$^{-1}$ varied
with the integration time on source, while the lower panel shows the
relative error deviate from the radiometer equation. The radiometer
equation, also the limiting sensitivity of the spectrometer, is
given by
$$\frac{\Delta T_{\rm rms}}{T_{\rm sys}} = \frac{K}{\sqrt{\Delta \nu \tau}} \, , \eqno(A1)$$
where $T_{\rm sys}$ is system temperature, $\tau$ is the sum of the
integration time on the source and on the off position, $K$ is a
factor accounting for data taking procedures, and $\Delta T_{\rm
rms}$ is rms noise temperature of the spectra for a given frequency
resolution $\Delta \nu$.

Figure~\ref{Fig7}a and b show the data obtained from the observing
semesters of 2008 to 2009 and of 2009 to 2010, respectively. It can
be seen in Figure~\ref{Fig7}a that the Allan plot begin to deviate from the
radiometer equation when integrating about 10 minutes on source, and
the relative error increase to 80\% when integrating 100 minutes,
however, it is shown in Figure~\ref{Fig7}b that the relative error
increase to only 10\% with the same integration time. Therefore,
both the stability and the sensitivity of the telescope have been
greatly improved after the system maintenance in the summer of 2009.
The average noise level in units of \ $T^*_A$ would be 0.020 K and
0.018 K when integrating 100 and 200 minutes on source respectively,
with spectra velocity resolution of $\sim$10\,km s$^{-1}$.
Therefore, this plot can be used to estimate observation time
according to the sensitivity that we required. But an important
thing to note in this Allan Variance Plot is that the data we used
are the raw data, that is to say, we didn't do any data processing
such as rejecting the spectra with distorted baseline or abnormal
rms noise level. Consequently, we could get even better sensitivity
if we just co-added the spectra that found to be consistent within
normal rms noise level. Based on these analysis, it seems to imply
that the telescope still have the capability to detect even weaker
emission signal than what have been detected, since the effective
integration time of most of our observations have not reach the
limit.

\end{document}